\documentclass[aps,prd,preprintnumbers,nofootinbib,twocolumn,superscriptaddress]{revtex4-2}

\usepackage[english]{babel} 
\usepackage[T1]{fontenc}
\usepackage{float}
\usepackage{setspace}
\usepackage{appendix}
\usepackage{parcolumns}
\usepackage{times}
\usepackage{type1cm}
\usepackage{url}
\usepackage{subfigure}
\usepackage{braket}
\usepackage{graphicx}
\usepackage{sidecap}
\usepackage{xspace}
\usepackage{tikz}
\usepackage{indentfirst}
\usepackage[mathscr]{eucal}
\usepackage{geometry}
\usepackage{booktabs}
\usepackage{indentfirst}
\usepackage{bm}
\usepackage{multirow}
\usepackage{dcolumn}
\usepackage{graphicx}
\usepackage{mathrsfs}  
\usepackage{enumerate}
\usepackage[utf8]{inputenc}
\usepackage{amsmath}
\usepackage{amssymb}
\usepackage{amsfonts}
\usepackage{amsthm}
\usepackage{mathrsfs}
\usepackage{graphicx}
\newcommand{\widesim}[2][1.5]{
	\mathrel{\underset{#2}{\scalebox{#1}[1]{$\sim$}}}
}

\usepackage{amsmath,mathtools}
\usepackage{comment}
\usepackage{amsbsy}
\usepackage{hyperref}

\makeatletter
\@addtoreset{equation}{section}

\makeatother

\def\Tr{{\rm Tr}}

\newcommand{\be}{\begin{equation}}
\newcommand{\ee}{\end{equation}}
\newcommand{\bea}{\begin{eqnarray}}
\newcommand{\eea}{\end{eqnarray}}

\newcommand{\nn}{\nonumber\\}

\newcommand{\Eqref}[1]{Eq.~(\ref{#1})}
\newcommand{\MSbar}{$\overline{\text{MS}}\,$}
\begin{document}

\title{Limit of vanishing regulator in the functional renormalization group}

	\author{Alessio Baldazzi}
\email{abaldazz@sissa.it}
\affiliation{International School for Advanced Studies, via Bonomea 265, I-34136 Trieste, Italy and INFN, Sezione di Trieste, Italy}

\author{Roberto Percacci}
\email{percacci@sissa.it}
\affiliation{International School for Advanced Studies, via Bonomea 265, I-34136 Trieste, Italy and INFN, Sezione di Trieste, Italy}

\author{Luca Zambelli}
\email{luca.zambelli@bo.infn.it}
\affiliation{\mbox{\it INFN-Sezione di Bologna, via Irnerio 46, I-40126 Bologna, Italy}}


\begin{abstract}
The non-perturbative functional renormalization group equation 
depends on the choice of a regulator function,
whose main properties are a ``coarse-graining scale'' $k$
and an overall dimensionless amplitude $a$.
In this paper we shall discuss the limit $a\to0$ with $k$ fixed.
This limit is closely related to the pseudo-regulator that reproduces
the beta functions of the \MSbar scheme,
that we studied in a previous paper.
It is not suitable for precision calculations
but it appears to be useful to eliminate
the spurious breaking of symmetries by the regulator,
both for nonlinear models and within the background
field method.

\end{abstract}

\maketitle

\section{Introduction}
\label{sec:intro}

\begin{figure}[H]
	\centering
	\includegraphics[width=.97\columnwidth]{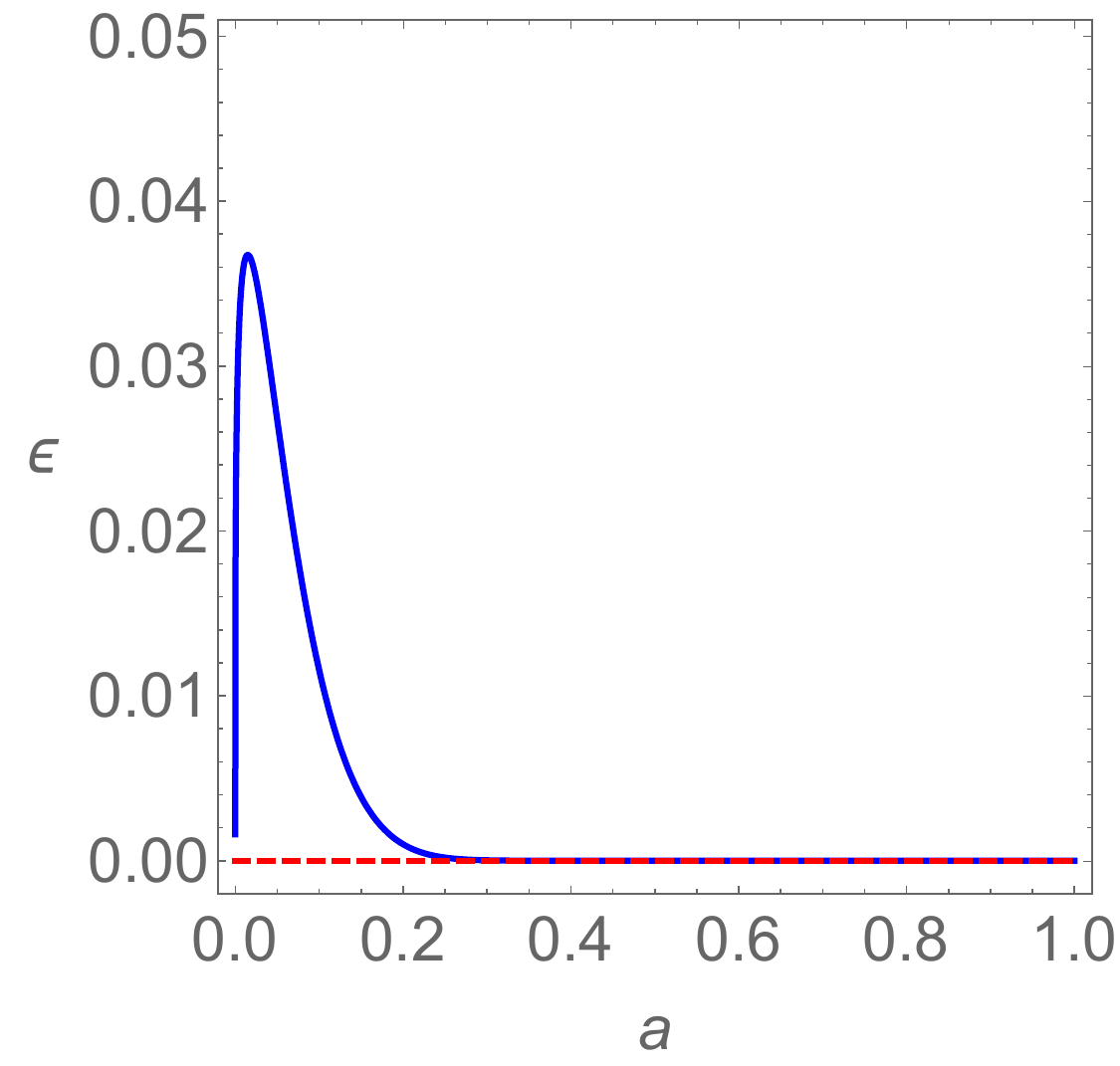}
	\caption{Blue, continuous curve: a path that reproduces the beta functions of dimensional regularization. Red, dashed curve:
		the limit of vanishing regulator.
		For a more detailed discussion see
		Sec.~\ref{sec:Ising}.}
	\label{fig:path}
\end{figure}

The Functional Renormalization Group (FRG) \cite{wett1,Morris:1993qb,Bonini:1992vh,Ellwanger:1993mw}
is a powerful tool to study 
quantum and statistical field theories and their
applications in statistical mechanics, condensed matter theory,
high energy physics \cite{Dupuis:2020fhh}.
It describes a continuous interpolation between an UV action
describing some microscopic physics and the Effective Action (EA)
where all the quantum/statistical fluctuations have been integrated out.
The functional that provides this interpolation is called the
Effective Average Action (EAA) and is denoted $\Gamma_k[\phi]$,
where $\phi$ are the fields, $k$ is a coarse-graining scale and 
$\Gamma_0=\Gamma$ is the EA.
The EAA can be defined by a functional integral with a cutoff
suppressing the contribution of low-momentum modes,
thus realizing Wilson's idea of integrating out high momentum modes first.
The cutoff itself is implemented by adding to the action the term
\be
\Delta S_k[\phi]=\frac12\int d^dx\,\phi R_k(-\partial^2)\phi\ ,
\label{cutoff}
\ee
leading to the functional differential equation
\be
k\frac{d\Gamma_k}{dk} = \frac{1}{2} \Tr \left( \frac{\delta^2 \Gamma_k}{\delta \phi \delta \phi} + R_k \right)^{-1} k\frac{d R_k}{dk}\ .
\label{FRGE}
\ee
This provides a non-perturbative definition of RG
that reduces to the perturbative one in the appropriate
domain \cite{Papenbrock:1994kf,Bonini:1996bk,Bonanno:1997dj,Pernici:1998tp,Kopietz:2000bh,Zappala:2002nx,Arnone:2003pa,Codello:2013bra}.
In this context, comparison with the results of
dimensional regularization become meaningful.
In Ref.~\cite{BPZ} we have discussed a two-parameter
family of regulators $R_k(a,\epsilon)$ that includes 
(for $\epsilon=0$)
a popular class of regulators used in the FRG literature.
On the other hand, taking the limit $a\to 0$
and $\epsilon\to 0$ (in this order),
it reproduces the beta functions of \MSbar.
In this paper we shall discuss what happens
when the limits are taken in the opposite order
(see Fig.~\ref{fig:path}).

We next discuss the motivation for this study.
The notation $\Gamma_k[\phi]$ emphasizes the important dependence of
this functional on the scale $k$,
but $\Gamma_k$ also depends on the shape of the cutoff function
$R_k$.
The notation $\Gamma[\phi,R_k]$ would thus be more appropriate,
and (\ref{FRGE}) could be replaced by a functional equation where
the derivatives with respect to $k$ are replaced by functional derivatives
with respect to $R_k$.
As mentioned above, all the relevant physical information is contained
in the EA and therefore {\it a priori},
all the dependence on $R_k$ is unphysical,
including the dependence on $k$.
However, there are situations where $k$ can be identified with
a physical parameter that acts in the theory as an IR cutoff.
In these cases, the dependence on $k$ can assume a physical
meaning.

Even though in such cases the dependence on $k$ reproduces the
dependence on physical parameters, 
the dependence on the shape of $R_k$ still remains unphysical.
Thus any observable must be independent of this shape.
On the other hand, when one makes approximations,
even physical observables will exhibit some spurious dependence on
the shape of the cutoff.
We will refer to this as ``cutoff dependence''.
\footnote{It is distinct from, 
but closely related to the ``scheme dependence''
of renormalized perturbation theory.}
For example, in statistical physics,
the position of a fixed point is not universal,
but the critical exponents are.
Still, when one calculates the critical exponents,
one must use some approximation and the results always depend
on the shape of the regulator.
In a specific calculation, one can then try to exploit 
this cutoff dependence to optimize the cutoff,
{\it i.e.} to find the cutoff that yields the best possible
value for the observables.
This is, in practice, an important aspect of FRG studies
\cite{Litim:2001up,Litim:2001fd,Litim:2002cf,Canet:2003qd,Balog:2019rrg,DePolsi:2020pjk}.

The main motivation for this study comes from another issue that
arises in certain applications of the FRG.
The central idea is simple and can be stated in great generality.
Suppose that the action at the microscopic level is invariant
under certain transformations.
Since the symmetry reflects physical properties of the system,
one would like to maintain it in the course of the RG flow.
However, for technical reasons,
it may be difficult to construct a regulator that has the symmetry,
and in this case the EAA will not have it either.
To be more precise, the classical symmetry of the  bare action
is translated into a ``quantum'' symmetry of the EAA,
which is deformed by the presence of the regulator. 
The latter symmetry is only implicitly determined,
as the corresponding regulator-dependent Ward identity
cannot in general be analytically and exactly solved~\cite{Ellwanger:1994iz} .
This will give rise to unpleasant complications.
Intuitively, we may try to minimize the breaking of the symmetry
by making the regulator as ``small'' as possible.
Let us make this notion a bit more precise.
For dimensional  reasons, we can write the regulator
as 
\begin{equation}
	R_k(z)=k^2 r_a(y)=k^2 a r_1(y) 
	\label{eq:generaldefofa}
\end{equation}
and $r_1$ is a dimensionless function of the dimensionless variable $y=z/k^2$, that is assumed to satisfy
the normalization condition $r_1(0)=1$
and $a$ is a positive real number.
\footnote{
Consider a fixed shape function $r_1$, such that $r_1(y)=0$ for $y>1$.
The limit $a\to\infty$ 
is expected  to completely remove from the path integral all the fluctuations
with momenta $q^2<k^2$. 
This is often referred to 
as the sharp cutoff limit.
Numerically optimal results are usually
obtained for $a\approx 1$.}
In many applications it is convenient 
to choose a shape function $r_1$ depending on 
some of the parameters appearing in the
ansatz adopted for the EAA.
The most common example is the insertion of an
overall wave-function-renormalization factor $Z_k$.
In this paper we 
shall mainly  
neglect these subtleties, as in most of our
	studies we will truncate the effective action
	to a scale-dependent local effective potential, and will
 be concerned with the limit $a\to 0$,
which we call the limit of vanishing cutoff.
\footnote{Thus, vanishing cutoff should not be misinterpreted as
$k\to 0$.}
One expects that in this limit the spurious effects
due to the breaking of the symmetry by the regulator
can be removed, or at least minimized.
It may seem that this limit is trivial, because for $a=0$
there is no cutoff, and the right-hand-side (rhs) of the exact FRG equation vanishes,
but we shall see that some important physical information 
remains available even in this limit.

Even though many of the challenges and properties
of the vanishing-regulator limit can be expected to characterize
large families of shape functions $r_1$, 
in this paper we
mainly focus on the following regulator choice	
\be
R_k(z) = a(k^2-z) \theta(k^2-z) \,,
\label{litima}
\ee
as in several interesting cases it is 
hardly feasible to study the
vanishing regulator limit
without having first specified 
a shape function.
The reasons for this are explained in Sec.~\ref{sec:vanishingVScs}
and further discussed in Sec.~\ref{sec:discussion}.


In order to better explain the problems
arising from the use of vanishing regulators,
and ways to circumvent them,
it is best to focus on simple and well-understood systems.
In Sec.~\ref{sec:oscill} we consider the harmonic and anharmonic oscillator.
Some of the features of vanishing regulators appear already in these cases.
In Sec.~\ref{sec:Ising} we deal with 
the  $\mathbb{Z}_2$--invariant scalar field
theory in $d\geq2$ Euclidean
dimensions, and its RG fixed point
(representing the Ising universality class).
We find that the main features of the Wilson-Fisher (WF) fixed point remain
accessible in the limit of vanishing regulator,
but the best approximation 
(after this limit is taken and among all possible
polynomial truncations of the potential)
 for the correlation-length critical exponent $\nu$ 
is obtained with the simplest
truncation, that only involves relevant couplings
(the mass and the quartic coupling).
There we also discuss the relation between the vanishing-$a$
limit of \eqref{litima} and the constant (momentum-independent) 
regulator, as well as the subtleties
concerning the application of vanishing regulators in 
an even number of dimensions.

In Sec.~\ref{sec:ON+1} we address the $O(N+1)$ nonlinear sigma model,
using a particular coordinate system on the sphere $S^N$.
This is an example of a system where the regulator breaks the
symmetry of the theory (respecting only the subgroup $O(N)$)
but in the limit of vanishing regulators the symmetry is seen
to be restored.
In Sec.~\ref{sec:background} we discuss a similar problem
that arises in applications of the background field method.
It is generally the case that the regulator breaks the
symmetry of the classical action consisting of
equal and opposite shifts of the background and 
fluctuation fields.
Also this symmetry is seen to be restored in the limit
of vanishing regulators.
%
%
We conclude in Sec.~\ref{sec:discussion} with a brief discussion of our results and some outlooks.
Some auxiliary formulas and analyses are provided in two appendices.

\section{Quantum oscillators}
\label{sec:oscill}

In this section we shall consider a very simple application of
the FRG equation as a tool to compute the EA at $k=0$.
This will allow us to investigate the effect of the
vanishing regulators on the calculation of 
some physical observable.
We shall consider first the simple harmonic oscillator
and then the anharmonic one.

The general bare action we are interested in reads
\be
S = \int dt \left( \frac{1}{2}\dot x^2 + \frac{1}{2}\omega^2 x^2 + \frac{\lambda}{4!}x^4 \right) \ .
\label{osc_bare}
\ee
In the Local Potential Approximation (LPA),
which is the first term in a derivative expansion,
the EAA is approximated by
\be
\Gamma_k = \int dt \left( \frac{1}{2}\dot x^2 + V_k(x) \right) \ .
\label{eq:LPAx}
\ee
Using the regulator in \Eqref{litima}
we get the following flow equation for the potential
\begin{align}
\partial_k V_k =&\ \frac{1}{\pi} \Bigg( \frac{a \, k \arctan \left(k \sqrt{\frac{1-a}{a k^2 + V''_k}}\right)}{\sqrt{(1-a) \left(a k^2 + V''_k \right)}}\nonumber\\
&- \sqrt{\frac{a}{ 1-a }} \arctan \left( \sqrt{\frac{1-a}{a}}\right)\Bigg) \,.
\label{osc_LPA}
\end{align}
The second term on the rhs
is equal to the first term evaluated at $V_k=0$.
This subtraction is not \textit{ad hoc}, and
is actually always meant to be present in the FRG
equation \cite{Lippoldt:2018wvi}, 
although in most applications it is
dropped since it only affects the 
ground-state energy.~\footnote{The
correct counterpart of
this term in applications to gravity,
especially within the 
background-field formulation, 
is still uncertain.} 
This term is due to the regularization of the functional measure of the
path integral, which ensures complete
suppression of the functional integral
in the $k\to\infty$ limit. 
As it is uniquely defined
to reproduce the Weyl ordering prescription,
it gives rise to the conventional ground-state energies
of the non-gravitating quantum/statistical mechanical models \cite{Gies:2006wv,Vacca:2011fx}.

Expanding the potential into a Taylor series
\be
V_k(x) = E_k + \frac{1}{2}\omega_k^2 x^2 + \frac{\lambda_k}{4!}x^4 + \ldots \ ,
\label{Vexp}
\ee
the beta functions are
%
%
%
\begin{subequations}
\begin{align}
 k\partial_k E_k =&\
\frac{k}{\pi} \Bigg[ \frac{a \,\arctan 
\left(
\sqrt{\frac{1-a}{a+\tilde\omega_k^2}}\right)}{\sqrt{(1-a)\left(a+ \tilde\omega_k^2\right)}}
\nonumber\\
& - \sqrt{\frac{a}{ 1-a }} \arctan \left( \sqrt{\frac{1-a}{a}}\right)\Bigg]
\ ,
\label{eq:flowEk}\\
 k\partial_k \omega_k^2 =& \
-\frac{a\lambda_k}{2\pi k} 
\frac{\frac{\sqrt{1-a}\sqrt{a+\tilde\omega_k^2}}{1+\tilde\omega_k^2}
+ 
\arctan\left(\sqrt{\frac{1-a}{a+\tilde\omega_k^2}}\right)}{(a+\tilde\omega_k^2)^{3/2}\sqrt{1-a}} ,
\label{eq:flowomegak}\\
 k\partial_k \lambda_k =& \
\frac{3a\lambda_k^2}{4\pi k^3} \Bigg[
\frac{3+2a+5\tilde\omega_k^2}{(1+\tilde\omega_k^2)^2
(a+\tilde\omega_k^2)^{2}}
\nonumber\\
& +\frac{
	3\arctan\left(\sqrt{\frac{1-a}{a+\tilde\omega_k^2}}\right)}{(a+\tilde\omega_k^2)^{5/2}\sqrt{1-a}}\Bigg]
\ ,
\label{eq:flowlambdak}
\end{align}
\end{subequations}
where $\tilde\omega_k=\omega_k/k$.

\subsection{Harmonic oscillator}
\label{sec:HO}

\begin{figure}[H]
	\begin{center}
		\includegraphics[width=0.97\columnwidth]{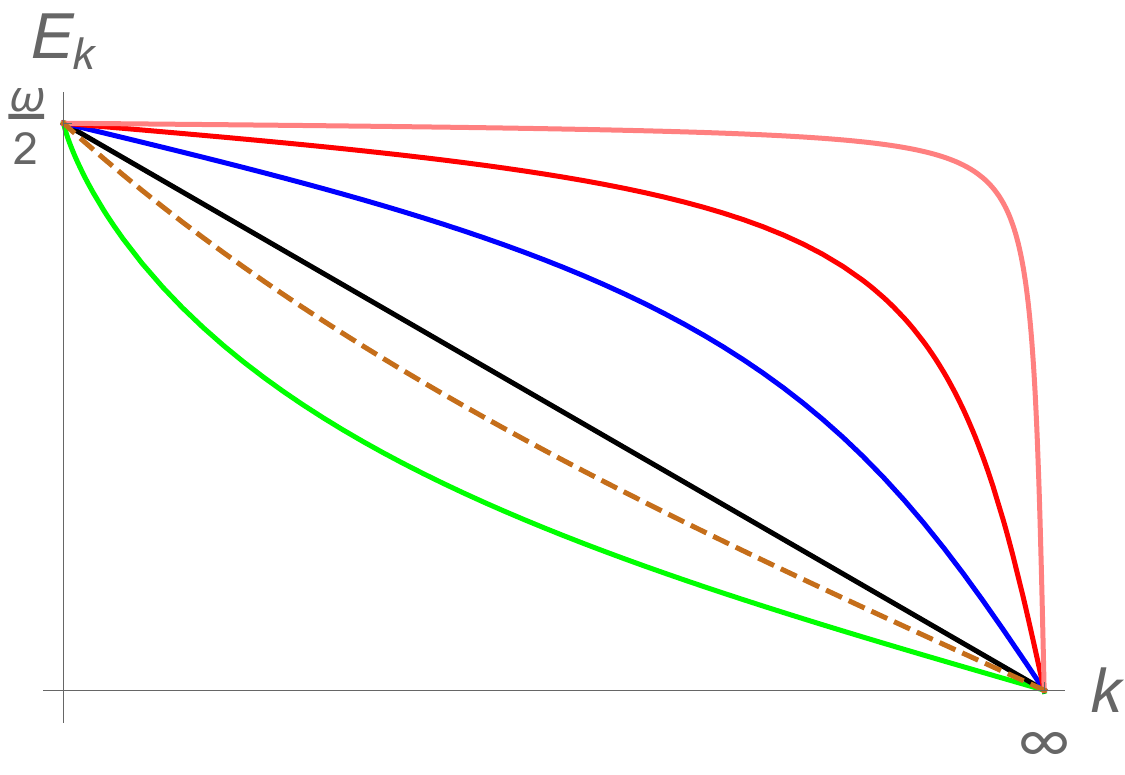}
		\caption{The continuous curves are
			the RG trajectories (\ref{ek}) for the harmonic
			oscillator ground-state energy $E_k$ for 
			the regulator (\ref{litima}) and various values of $a$.
			From bottom to top: $a\to\infty$ (sharp cutoff, green),
			$a=1$ (black), $a=1/10$ (blue), $a=1/100$ (red),
			$a=1/10000$ (pink).
			The dashed curve is the flow (\ref{ekcs}).
			The horizontal axis has been rescaled by the function $k=\tan(\pi x/2)$.}
		\label{fig:trajectoriesa}
	\end{center}
\end{figure}

We start by addressing the computation of the vacuum
energy of the harmonic oscillator.
When $\lambda_k=0$, Eq.~(\ref{eq:flowomegak}) shows that
$\omega_k$ is independent of $k$,
thus we shall simply write $\omega_k=\omega$.
The solution of (\ref{eq:flowEk}) is
\begin{align}
&E_k = \frac{\omega}{2}
+\frac{k}{\pi} 
\Bigg[
\sqrt{\frac{a+\tilde\omega^2}{1-a}}
\arctan\left(\sqrt{\frac{1-a}{a+\tilde\omega^2}}\right)
\nonumber\\
&-
\sqrt{\frac{a}{1-a}}
	\arctan
\left(\sqrt{\frac{1-a}{a}}\right)
-\tilde\omega  
\arctan\left(\frac{1}{\tilde\omega}\right)\Bigg] \,.
\label{ek}
\end{align}
This function is plotted in Fig.~\ref{fig:trajectoriesa}
for various values of $a$.
First of all we see that $E_{0}=\omega/2$ for any $a$. 
The $a$-independence of the 
result is just an example of a more general 
phenomenon: while the $k$ dependence
of any quantity along an RG trajectory
is sensitive to the functional form
of $R_k$, the boundary values at $k\to+\infty$
and $k\to 0$ are not. 

The second point to notice is that the convergence of the
flow towards the IR becomes faster for decreasing $a$.
This can be understood as follows.
The regulator term is effective in suppressing the
propagation when it becomes comparable or larger than the
kinetic term, {\it i.e.} for $R_k(q^2)>q^2$.
For the regulator (\ref{litima}), this happens when
\be
q^2<k_\text{eff}^2\equiv \frac{a}{1+a}k^2\ .
\label{keff}
\ee
Thus decreasing $a$ has the same effect as decreasing $k_\text{eff}$.

From this discussion, there seems to be no issue with the limit
$a\to0$. 
Some subtlety appears however
when we try
to construct the RG trajectory from the $a\to0$ limit 
of \Eqref{eq:flowEk},
which reads
\be
k\partial_k E_k\sim
-\frac{k}{2}\sqrt{a}+O(a)\  .
\ee
This way of taking the limits is of course nonsensical,
as the resulting beta functions would be identically vanishing.
The $\omega$ independence is a consequence of
the fact that the numerator of the
rhs~of the FRG equation is already proportional to $a$.
However, we obtain a nontrivial equation if we rescale
\begin{align}
\omega^2 = a\,\hat{\omega}^2,\quad\quad
E_k=\sqrt{a}\ \hat{E}_k\ ,
\label{eq:arescmE}
\end{align}
and then take the $a\to0$ limit of \Eqref{eq:flowEk}.
This leads to the finite result
\be
k\partial_k \hat{E}_k=\frac{1}{2}\left(
\frac{k}{\sqrt{k^2+\hat{\omega}^2}}-1\right)+O(\sqrt{a})\ ,
\label{eq:flowEkCS}
\ee
which is the same flow equation one would find with 
a constant (often called a ``Callan-Symanzik'') regulator 
\begin{equation}
	R_k=k^2.
	\label{eq:CSreg}
\end{equation}
The latter leads to the flow
\be
E_k=\frac12\left(\sqrt{k^2+\omega^2}-k\right)\ ,
\label{ekcs}
\ee
which is plotted as the dashed curve in Fig.~\ref{fig:trajectoriesa}.

The rescaling (\ref{eq:arescmE}) is formulated
as an ``active'' transformation of the couplings $E_k$ 
and $\omega_k$ throught a factor $\sqrt{a}$, 
while the RG coordinate $k$ stays independent of $a$.
Within the flow equations
of the dimensionless couplings 
$\tilde{E}_k\equiv E_k/k$ and $\tilde{\omega}^2$,
it is also possible to reinterpret it 
as a ``passive'' transformation, in which
the dimensionful couplings $E_k$ and $\omega_k$
are independent of $a$, while the RG coordinate changes by
$k=\hat{k}/\sqrt{a}$.
This second interpretation is consistent with the observation
that $k_\mathrm{eff}\to \hat{k}$ for $a\to 0$ 
according to \Eqref{keff}. 


\subsection{Anharmonic oscillator}

Next, we turn to the anharmonic oscillator with $\lambda\neq0$.
As our interest is in comparing 
	different possible prescriptions for taking the $a\to0$ limit
with the available solutions for some interesting quantities,
we do not address the numerical analyses needed to
compute the energy levels, finding analytical expressions 
more instructive. We therefore
consider only the first order of the expansion in $\lambda$.
We see from \eqref{eq:flowlambdak} that at this order 
the beta function of $\lambda_k$ is zero.
Therefore $\lambda_k=\lambda$ at all scales.
Expanding the vacuum energy parameter to first order in $\lambda$
\be
E_k=E_k \Big|_{\lambda=0} + \frac{dE_k}{d\lambda}\Big|_{\lambda=0} \lambda +\ldots 
\ee
and solving the flow equation with the initial condition that
$\lim_{k\to\infty}\Gamma_k=S$ is the bare action (\ref{osc_bare}),
we find that the first order correction to the energy is
\begin{widetext}
\begin{align}
\frac{dE_k}{d\lambda}\Big|_{\lambda=0} =
&\ \frac{1}{32\omega^2}
+\frac{\arctan\left(\frac{1}{\tilde\omega }\right)
\left[2\tilde\omega \sqrt{\frac{1-a}{a+\tilde\omega^2}} 
\arctan\left(\sqrt{\frac{1-a}{a+\tilde\omega^2}}\right)
+(a-1)\left(\pi+\arctan\left(\frac{1}{\tilde\omega}\right)^2\right)\right]}{8\pi^2(a-1)\omega^2}
\nonumber\\
&
-\frac{\arctan\left(\sqrt{\frac{1-a}{a+\tilde\omega^2}}\right)
\left(\pi\sqrt{\frac{1-a}{a+\tilde\omega ^2}}
+\frac{\tilde\omega}{a+\tilde\omega^2}
\arctan\left(\sqrt{\frac{1-a}{a+\omega^2}}\right)\right)
}{8\pi^2(a-1)k\omega} \,.
\label{E1anharmonic}
\end{align}
\end{widetext}
On the other hand for the frequency we find to first
order in $\lambda$
\begin{align}
\omega_k^2 =&\ \omega^2 
+ \frac{\lambda}{4\omega}\Bigg[ 1 +
\frac{2\tilde\omega  \sqrt{\frac{1-a}{a+\tilde\omega^2}} 
\arctan\left(\sqrt{\frac{1-a}{a+\tilde\omega^2}}\right)}{\pi(1-a)}
\nonumber\\
&-\frac{2}{\pi}\arctan\left(\frac{1}{\tilde\omega}\right)\Bigg] \,.
\label{omegaanharmonic}
\end{align}
The quantities $\frac{dE_k}{d\lambda}\Big|_{\lambda=0}$ and $\omega_k$  given in
(\ref{E1anharmonic}-\ref{omegaanharmonic}) are the solutions of the flow equations
at arbitrary $k$. They interpolate between the initial conditions 
$\lim_{k\to\infty}\frac{dE_k}{d\lambda}\Big|_{\lambda=0}=0$ and 
$\lim_{k\to\infty}\omega_k=\omega$ and the corresponding parameters
in the EA at $k=0$.

Also in this case, it is not possible to directly take the limit $a\to0$
in the flow equations, because then the beta functions simply vanish.
However, having solved the flow equations
one can take the limits $a\to0$ and $k\to0$ in any order
obtaining
\begin{subequations}
\begin{align}
&\omega_0^2 = \omega^2 + \frac{\lambda}{4\omega} \,,
\\
&E_0 = \frac{\omega}{2} + \frac{\lambda}{32\omega^2} \,.
\end{align}
\label{charlie}
\end{subequations}

One can take the limit $a\to0$ in the flow equations
provided the potential is rescaled to:
\begin{subequations}
	\begin{align}
		V_k(x) &= \sqrt{a}\ \hat{V}_k(\hat{x})\ ,\\
		x &= a^{-1/4}\ \hat{x}\ .
	\end{align}
\end{subequations}
Expanding around $a=0$ the flow equation becomes
\be
\partial_k \hat{V}_k = \frac{1}{2} \left( \frac{k}{\sqrt{k^2+\hat{V}_k''}}-1 \right) \,.
\label{eq:subtractedVflow}
\ee
This is the flow equation of the potential
for a constant regulator.
Projecting the latter on a polynomial truncation of the potential
as in \Eqref{Vexp}, we deduce the beta functions
\begin{subequations}
\begin{align}
& \partial_k E_k = \frac{1}{2} \left( \frac{k}{\sqrt{k^2+\omega_k^2}}-1 \right) \ , 
\\
& \partial_k \omega_k^2 = -\frac{1}{4} \frac{k}{\left(k^2+\omega_k^2\right)^{3/2}}\lambda_k  \ ,
\\
& \partial_k \lambda_k = \frac{9}{8} \frac{k}{\left(k^2+\omega_k^2\right)^{5/2}}\lambda_k^2  \ .    
\end{align}
\end{subequations}
Solving these equations one reobtaines (\ref{charlie}).
\newline

\section{The Ising universality class}
\label{sec:Ising}

In this section we shall consider the theory of a single, 
$\mathbb{Z}_2$-invariant scalar field $\phi$ 
 in the LPA
\be
\Gamma_k=\int d^dx\left[\frac12(\partial\phi)^2
+V_k(\phi)\right]\ .
\label{LPA}
\ee
While the FRG equation allows us to treat the potential as a whole,
it will be instructive to further expand 
\be
V_k(\phi) = \sum_{n=0}^\infty \frac{\lambda_{2n}}{(2n)!} \phi^{2n}\ .
\label{potexp}
\ee
The term $n=0$ is the vacuum energy and can usually be ignored,
but we shall need it later in our discussion.
Then, from the FRG equation we can derive infinitely many beta functions
$\beta_{2n}=k\frac{\partial \lambda_{2n}}{\partial k}$.
For arbitrary regulator, and in any dimension,
for the first few couplings
this leads to
\begin{subequations}
\begin{align}
\beta_0 =&\
\frac{1}{2(4\pi)^{d/2}} Q_{d/2}\left[ \frac{\partial_t R_k}{P_k+\lambda_2} \right] \,,
\label{betaQ0}
\\
\beta_2 =&\
-\frac{1}{2(4\pi)^{d/2}} \lambda_4 Q_{d/2}\left[ \frac{\partial_t R_k}{(P_k+\lambda_2)^2} \right],
\label{betaQ2}
\\
\beta_4 =&\ \frac{1}{2(4\pi)^{d/2}}  \left(
6\lambda_4^2 Q_{d/2}\left[ \frac{\partial_t R_k}{(P_k+\lambda_2)^3} \right]\right.
\nonumber\\
&\left.-
\lambda_6 Q_{d/2}\left[ \frac{\partial_t R_k}{(P_k+\lambda_2)^2} \right]
\right),
\label{betaQ4}
\\
\beta_6 =&\ \frac{1}{2(4\pi)^{d/2}} \left(
-90 \lambda_4^3 Q_{d/2}\left[ \frac{\partial_t R_k}{(P_k+\lambda_2)^4} \right]
\right.
\nonumber\\
& 
+30 \lambda_4 \lambda_6 Q_{d/2}\left[ \frac{\partial_t R_k}{(P_k+\lambda_2)^3} \right]
\nonumber\\
& \left. -\lambda_8 Q_{d/2}\left[ \frac{\partial_t R_k}{(P_k+\lambda_2)^2} \right]
\right)\ ,
\label{betaQ6}
\end{align}
\label{eqs:betaQ}
\end{subequations}
where
\be
Q_n \left[ W \right]= \frac{1}{\Gamma(n)} \int_0^\infty dz \, z^{n-1} W(z)
\label{Qf}
\ee
are momentum integrals ($z={q}^2$).
These integrals can be evaluated in closed forms by using the
optimized regulator \eqref{litima}.
The $Q$ functionals are then given by hypergeometric functions
\begin{align}
Q_n\left(\frac{\partial_t R}{(P+m^2)^\ell}\right)
=&\ \frac{2ak^{2(n+1-\ell)}}{\Gamma(n+1)(a+\tilde m^2)^\ell}\nonumber\\
&\times\!\!\!\!\!
\phantom{A}_2F_1\left(\ell,n,1+n,\frac{a-1}{a+\tilde m^2}\right) .
\label{Qhyp}
\end{align}
These are plotted in Fig.~\ref{fig:Qa} for $d=3$ and $d=4$.
\begin{widetext}
	\begin{minipage}{\linewidth}
		\begin{figure}[H]
			\begin{center}
				\includegraphics[width=0.45\textwidth]{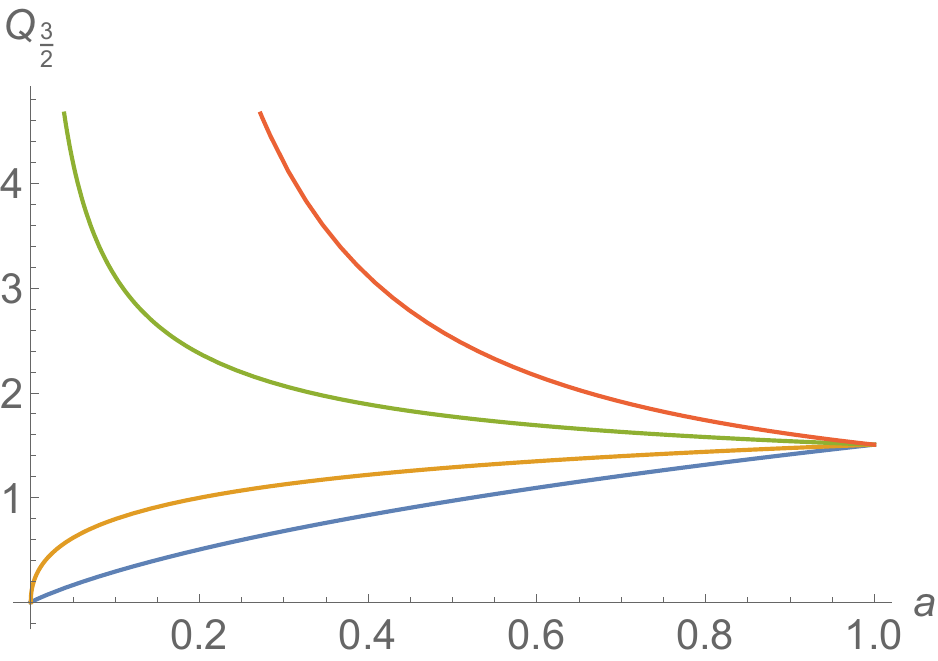}\quad 
				\includegraphics[width=0.45\textwidth]{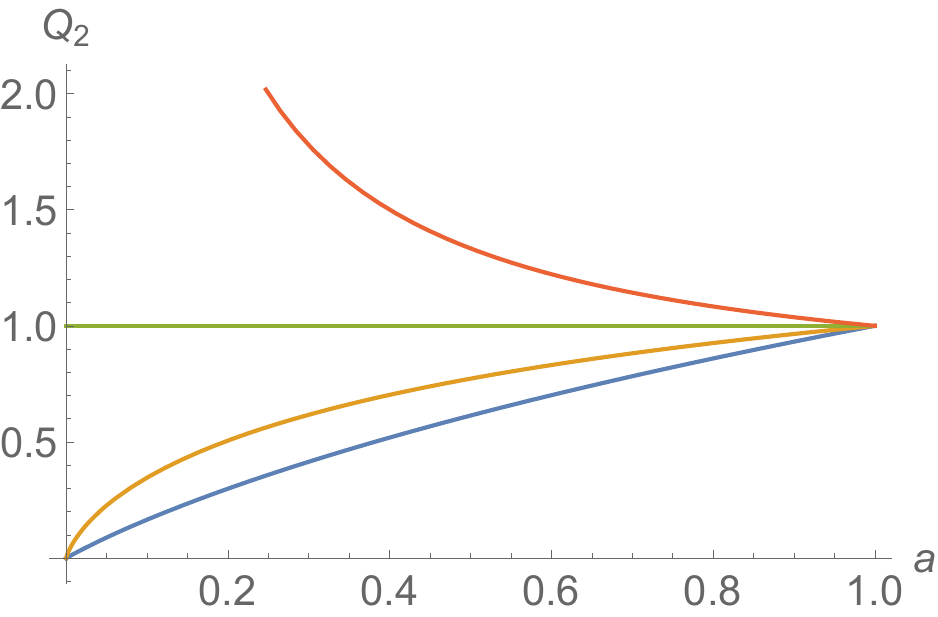}\\
				\includegraphics[width=0.45\textwidth]{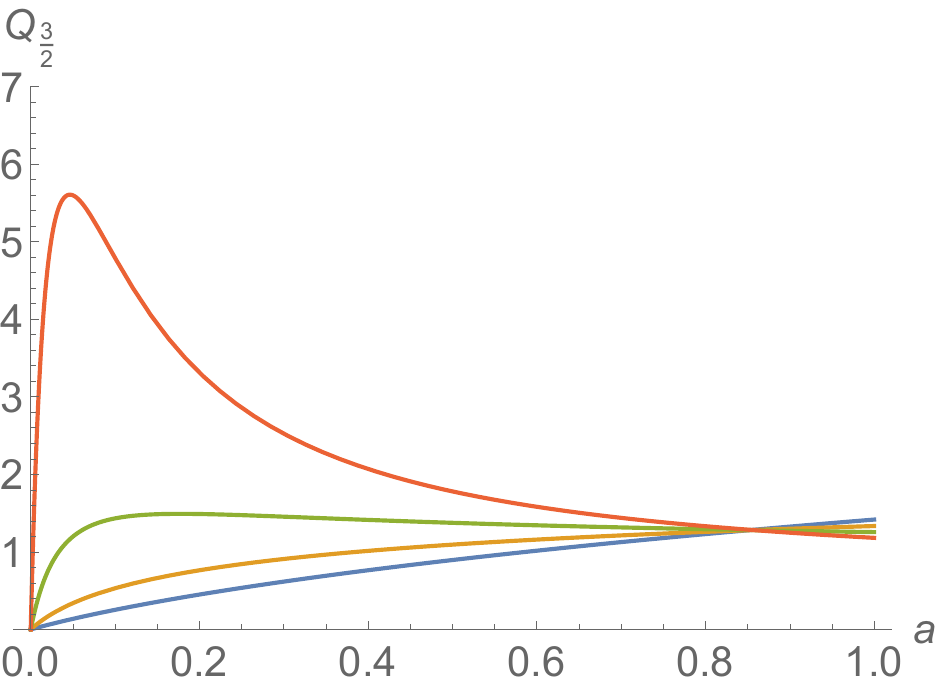}\quad 
				\includegraphics[width=0.45\textwidth]{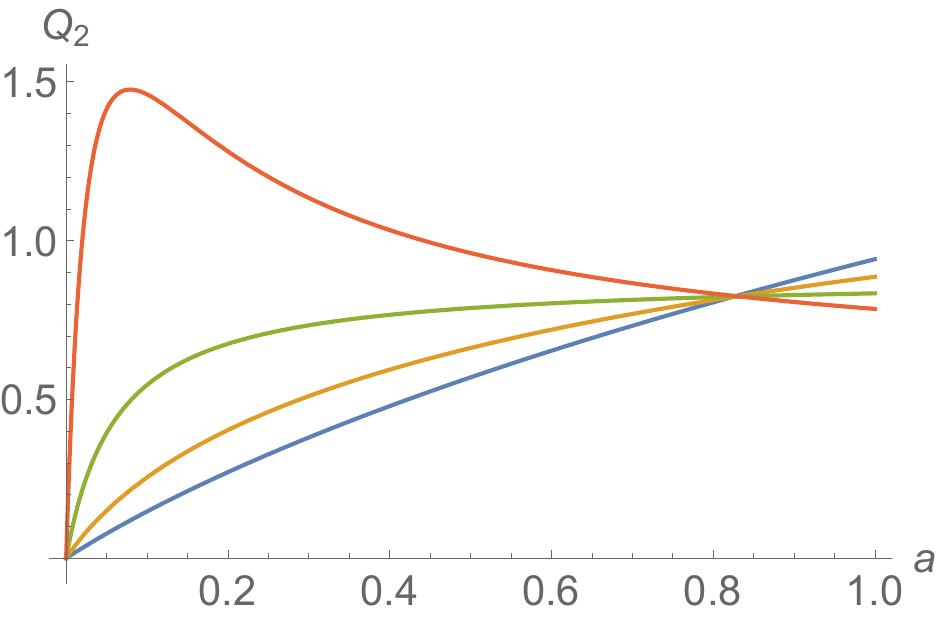}
				\caption{The functionals (\ref{Qhyp}) with $n=3/2$ ($d=3$, left)
					and $n=2$ ($d=4$, right), with $\tilde m=0$ (top line)
					and $\tilde m=0.25$ (bottom line).
					In each figure $\ell=1,2,3,4$, from bottom to top.}
				\label{fig:Qa}
			\end{center}
		\end{figure}
	\end{minipage}
\end{widetext}
In particular, in the massless case and
in the limit of 
vanishing regulator we obtain
\be
\lim_{a\to 0}Q_n\left(\frac{\partial_t R}{P^\ell}\right)
=\begin{cases}
0\quad\mathrm{for}\quad \ell<n+1\ ,\\
1\quad\mathrm{for}\quad \ell=n+1\ ,\\
\infty\quad\mathrm{for}\quad \ell>n+1\ .
\end{cases}
\ee
Clearly, the beta functions will not be finite.~\footnote{Note that these are infrared divergences:
in the massive case all $Q$ functionals go to zero for $a\to0$.}
For this reason an additional regularizing device
is needed to make sense of vanishing regulators.
In Ref.~\cite{BPZ} we have discussed a family of
regulators depending on an additional parameter $\epsilon$
that, in the limit $a\to0$ and $\epsilon\to 0$ (in this order)
reproduces the results of dimensional regularization.
In the $a$-$\epsilon$ plane the limit had to be taken
along a curve of the general form shown in Fig.~\ref{fig:path}.
In this paper we shall instead try to take the limits in the inverse order.
In fact, we shall not even talk about the parameter $\epsilon$
and try to take the limit $a\to0$ along the path $\epsilon=0$
(red, dashed) in Fig.~\ref{fig:path}.

Returning to the beta functions (\ref{eqs:betaQ}),
we note that if we set $\lambda_2=0$,
the beta functions of the relevant couplings go to zero,
those of the marginal couplings are independent of $a$
and those of the irrelevant couplings diverge in the
limit of vanishing regulator.
Given this rather singular behavior, one may fear that all
physical information gets lost in this limit.
Actually, this is not so,
as we intend to show in $d=3$, where the system is 
know to have
a nontrivial fixed point.

\subsection{The Wilson-Fisher fixed point: relevant couplings}
\label{WFLPArel}

In order to make our point it will be enough, 
as a first step,
to consider a truncation that contains only the relevant couplings
(we are now in $d=3$):
\be
V_k=\frac{\lambda_{2}}{2} \phi^2+\frac{\lambda_{4}}{24} \phi^4\ .
\label{tr24}
\ee
Defining the dimensionless variables 
\begin{equation}
 \tilde\lambda_{2n} = k^{-d+n(d-2)} \lambda_{2n} \,,
 \label{eq:dlesslambdas}
\end{equation}
the beta functions are
\begin{subequations}
\begin{align}
\tilde\beta_2 &= - 2 \tilde\lambda_2
-\frac{a \tilde\lambda_4 }{6 \pi^2 \left(a+ \tilde\lambda_2 \right)^2}
\,_2F_1\!\left( 2, \frac{3}{2}, \frac{5}{2}; \frac{a-1}{a+\tilde\lambda_2} \right),
\\
\tilde\beta_4 &= - \tilde\lambda_4 +\frac{a\tilde\lambda_4^2 }{\pi^2 \left(a+ \tilde\lambda_2 \right)^3}
\,_2F_1\! \left( 3, \frac{3}{2}, \frac{5}{2}; \frac{a-1}{a+\tilde\lambda_2} \right).
\end{align}
\label{ising24}
\end{subequations}
\vskip2mm
Expanding in $\tilde\lambda_2$
\begin{subequations}
\begin{align}
 \tilde\beta_2 =&\
  - 2 \tilde\lambda_2
-  \frac{a\tilde\lambda_4}{4\pi^2 (1-a)}\Bigg[ \Bigg(\frac{\arctan \left(\frac{\sqrt{1-a}}{\sqrt{a}}\right)}{\sqrt{(1-a) a}}-1\Bigg)
\nonumber\\
&\!\!\!\! +   \Bigg(\frac{2 a- 1}{2a} - \frac{\arctan \left(\frac{\sqrt{1-a}}{\sqrt{a}}\right)}{2a\sqrt{a(1-a) }}\Bigg)  \tilde\lambda_2 
\Bigg] ,
\end{align}
\begin{align}
\tilde \beta_4 =&\ -\tilde\lambda_4 + \frac{\tilde\lambda_4^2}{\pi^2 a^2}\, _2F_1\!\left(\frac{3}{2},3;\frac{5}{2};\frac{a-1}{a}\right)\! .
\end{align}
\end{subequations}
The WF fixed point is now located at
\begin{subequations}
\begin{widetext}
\be
\tilde\lambda_{2}^* =
\frac{2 a^3 \left(1-\frac{\arctan \left(\frac{\sqrt{1-a}}{\sqrt{a}}\right)}{\sqrt{(1-a) a}}\right)}{a^2 \left(2 a-1-\frac{\arctan \left(\frac{\sqrt{1-a}}{\sqrt{a}}\right)}{\sqrt{(1-a) a}}\right)+16 (1-a) \,
   _2F_1\left(\frac{3}{2},3;\frac{5}{2};\frac{a-1}{a}\right)} \ ,
\ee
\end{widetext}
\be
\tilde\lambda_{4}^* =
\frac{\pi^2 a^2}{ \, _2F_1\!\left(\frac{3}{2},3;\frac{5}{2};\frac{a-1}{a}\right)} \ .
\ee
\end{subequations}
If we expand the critical couplings around $a=0$
\begin{subequations}
\begin{align}
& \tilde\lambda_2^* \widesim[2]{a\to 0}
-\frac{2 a}{5}+o\left(a^{3/2}\right) \,,
\\
& \tilde\lambda_4^* \widesim[2]{a\to 0}
\frac{16 \pi  \sqrt{a}}{3}+ o(a^{3/2}) \,.
\end{align}
\end{subequations}
\newline
Thus the WF fixed point merges with the Gaussian fixed point.
Note that since $\tilde\lambda_2=\tilde m^2$ is linear in $a$
for $a\to0$ at the WF fixed point, the Q functional (\ref{Qhyp}) does not go to zero
and this entails that the  quantum/statistical
contribution to the critical exponents will be nontrivial for $a\to0$.

Indeed, the position of the fixed point is not physically significant.
If we consider the stability matrix at the nontrivial fixed point
\begin{widetext}
\be
M= \left( \frac{\partial\tilde\beta_i}{\partial\tilde\lambda_j} \right)_*
=\left(
\begin{array}{cc}
-\frac{5}{3}
 & 
\frac{4 a \, _2F_1\!\left(\frac{3}{2},3;\frac{5}{2};\frac{a-1}{a}\right) \left(1-\frac{\tan ^{-1}\left(\frac{\sqrt{1-a}}{\sqrt{a}}\right)}{\sqrt{(1-a) a}}\right)}{\pi^2 a^2 \left(2 a-\frac{\tan
   ^{-1}\left(\frac{\sqrt{1-a}}{\sqrt{a}}\right)}{\sqrt{(1-a) a}}-1\right)+16 (1-a) \, _2F_1\!\left(\frac{3}{2},3;\frac{5}{2};\frac{a-1}{a}\right)}
 \\
0 & 1 \\
\end{array}
\right) 
\ee
\end{widetext}
we see that the component $(1,2)$ of $M$ goes to zero for $a \to 0$ and so the stability matrix becomes diagonal.
The eigenvalues of $M$, that is minus the critical exponents $\theta_i$,
are actually independent of $a$, 
in particular $\nu=(\theta_1)^{-1} = 0.6$.
We see that even though the WF fixed point collapses 
towards the Gaussian one, it keeps its distinct character
in the limit $a\to0$ and a different critical exponent $\nu$.
In fact, the numerical value is not very bad,
considering the drastic approximation.

\subsection{The Wilson-Fisher fixed point in the LPA}
\label{WFLPAvana}

\vskip-4mm
Let us now treat the potential as a whole \cite{Morris:1994ie}.
Inserting (\ref{LPA}) in the FRGE we obtain the ``beta functional''
\be
\partial_t V_k = \frac{1}{2(4\pi)^{d/2}} Q_{d/2} \left[ \frac{\partial_t R_k}{P_k+ V_k'' } \right]\ .
\label{betaV}
\ee
Using the regulator (\ref{litima}), setting $d=3$
and rescaling
\begin{subequations}
\begin{align}
	\phi &=\frac{k^{1/2}}{\pi\sqrt{6}} \tilde{\phi}\ ,
\end{align}
\begin{align}
	V_k(\phi) &=\frac{k^3}{6\pi^2} v(\tilde{\phi})\ ,
\end{align}
\end{subequations}	
the beta function of the dimensionless potential
$v$
becomes
\be
\partial_t v = - 3 v +\frac{1}{2}\tilde{\phi} v'  +\frac{a}{a+ v''}\ {}_2F_1\! \left(1,\frac{3}{2},\frac{5}{2}; \frac{a-1}{a+ v''}\right) \,.
\label{eq:flowPLAa3d}
\ee
We look for even scaling solutions shooting from the origin 
with initial condition $v''(0)$ and $v'(0)=0$.
There are only two values of $v''(0)$ which can be identified as fixed-point solutions: $v''(0)=0$, that corresponds to
the Gaussian fixed point, and some negative value
that corresponds to the WF fixed point.
As in the preceding section,
for decreasing values of $a$,
the WF fixed point moves towards the Gaussian one.
We see that also in the functional treatment,
the WF fixed point collapses into the Gaussian one.

This is confirmed by shooting from infinity.
The potential for the WF solution has the following asymptotic behavior for large field
\begin{align}
&v= A \tilde{\phi}^6+a \Bigg(\frac{1}{150 A \tilde{\phi} ^4}-\frac{2 a+3 }{31500 A^2 \tilde{\phi} ^8}
\nonumber\\
&
+\!\frac{8 a^2+12 a+15}{8505000 A^3 \tilde{\phi} ^{12}}-\frac{a}{67500 A^3 \tilde{\phi} ^{14}}+O\!\left( A^{-4} \tilde{\phi} ^{-16}\right)\!\! \Bigg).
\end{align}
The free parameter $A$ can be fixed as function of $a$ by requiring $\mathbb{Z}_2$ symmetry for vanishing field \cite{Bridle:2013sra}.
We find that in the limit $a \to 0$, $A$ tends to 
$A\approx 0.0015$.~\footnote{ The asymptotic parameter  is $A=0.001$ for $a=1$ and it increases monotonically for $a \to 0$.}

\begin{figure}[H]
	\begin{center}
		\includegraphics[width=0.97\columnwidth]{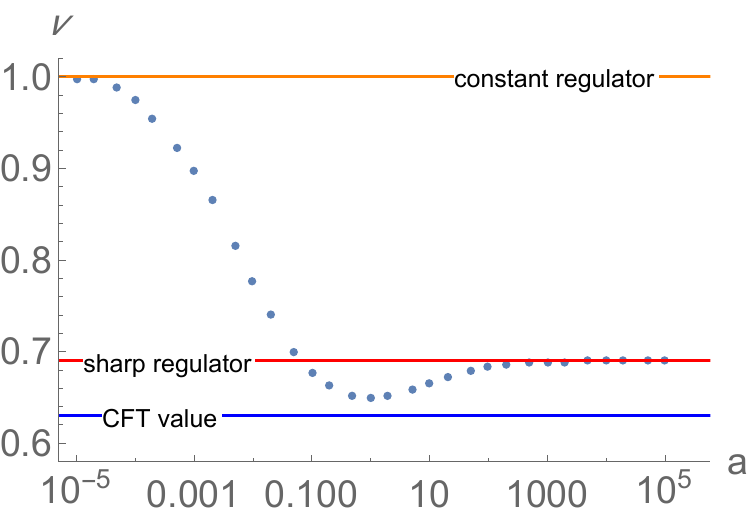}
		\caption{The dots represent the values of the critical exponent 
			$\nu$ as a function of $a$. For comparison we have drawn the
			values of $\nu$ for the sharp regulator and the constant
			(mass) regulator,
			as well as the 
			conformal-bootstrap value~\cite{Kos:2014bka}. This figure extends Fig.~12 in Ref.~\cite{Litim:2002cf}
			to low values of $a$.}
		\label{fig:nuevareg}
	\end{center}
\end{figure}

The scaling exponents $\theta_i$ are obtained by 
linearizing the flow equation around the fixed point
and calculating the spectrum of eigenperturbations.
The analysis has to be done numerically.
For the Gaussian fixed point the spectrum
is independent of $a$.
Figure~\ref{fig:nuevareg} gives $\nu$ of the WF fixed point
as a function of $a$ for $10^{-5}<a<10^5$.
As expected, the best value is obtained for $a\approx 1$,
while in the limit of vanishing regulator $\nu$
appears to approach $\nu=1$.
Besides the correlation-length exponent $\nu = (\theta_1)^{-1}$, we also find positive eigenvalues, as reported in Tab.~\ref{tab:critexpWFa10}.

For vanishing $a$
all the scaling exponents are odd integers.
This coincides with
the spectrum of 
the $O(N)$ model in the limit of large $N$,
which is known, and we have indeed checked,
to be independent of $a$~\cite{DAttanasio:1997yph,Marchais:2017jqc}.

\begin{table}[H]
	\begin{center}
		\begin{tabular}{ |c|c|c|c|c|c| } 
			\hline
			& $\theta_1$ &  $\theta_2$ & $\theta_3$ & $\theta_4$ & $\theta_5$ \\ 
			\hline
			$a=1$ & $1.539$ &  $-0.656$  &  $-3.180$ & $-5.912$ & $-8.796$ \\ 
			$a=0$ & $1$ & $-1$ & $-3$ & $-5$ & $-7$ \\
			\hline
		\end{tabular}
	\end{center}
	\caption{The first few critical exponents
		at the Wilson-Fisher fixed,
		point computed
		in the local potential approximation for the 
		regulator~\eqref{litima}.
		We report the most common
		choice $a=1$ and the
		limiting case of the vanishing
		regulator.}
	\label{tab:critexpWFa10}
\end{table}



\subsection{Vanishing regulators and constant regulators}
\label{sec:vanishingVScs}

At this point it is relevant to recall
that the critical exponent $\nu=1$
is known to result also from
the LPA equations for a
constant regulator \eqref{eq:CSreg}~\cite{Litim:2002cf}.
 Together with the findings of 
Sec.~\ref{sec:oscill}, this observation
points at a more general result which we detail
in this section.

So far  we have
first solved the fixed point equations
for generic $a$ and then sent $a\to 0$.
On the other hand, we are now going to argue that
when the vanishing-regulator limit is taken
on the LPA beta functions,
i.e.~before integrating the flow,
it results, in general non-even $d$,
in the flow equations of the constant regulator.

The first way of reaching
this conclusion is based on a redefinition
of the RG scale $k$, which we have already
introduced in Sec.~\ref{sec:oscill}.
Suppose that in addition to the parameter $a$ we also
introduce a parameter $b$ rescaling the cutoff $k$
\be
R_k(z) = a\left(b k^2-z \right) \theta \left(b k^2 -z \right) \,.
\ee
This rescaling can be motivated as follows.
First of all, it should not change the scaling solutions.
Furthermore, as discussed in Sec.~\ref{sec:HO}, 
we can define an ``effective'' cutoff scale $k_\text{eff}$
as the momentum scale where the cutoff term $R_k$ 
becomes comparable to the kinetic term.
If we decrease $a$, the effective cutoff scale also decreases.
It was suggested in Ref.~\cite{Litim:2002cf} that the decrease of $a$
should be compensated by choosing $b$ so that at some
conventional scale $z_0<k^2$,
the regulator is normalized: $R_k(z_0)= k^2$.
This fixes $b=\frac{1}{a}+\frac{z_0}{k^2}$,
leading to the regulator
\be
R_k(z) =  \left(k^2-a\left( z-z_0 \right) \right) \theta \left( k^2 -a\left( z-z_0 \right) \right)\,.
\label{eq:rescaledlitim}
\ee
Now we see that in the limit $a \to 0$,
the regulator becomes a constant
as in \Eqref{eq:CSreg}.
The latter leads to the dimensionless flow equation
\be
\partial_t v = -d v +\left(\frac{d}{2}-1 \right) \tilde{\phi} v' 
+ \frac{\pi \left(1+v''\right)^{\frac{d}{2}-1}}{(4\pi)^{d/2}\Gamma \left(\frac{d}{2}\right) \sin \left( \frac{d\pi}{2}\right)}  \ .
\label{cs}
\ee
In $d=3$ and after the rescaling $v \to v/(4\pi)$ and $\tilde{\phi} \to \tilde{\phi}/\sqrt{4\pi}$
this takes the simple form
\be
\partial_t v = - 3 v +  \frac{1}{2} \tilde{\phi} v' -  \sqrt{1+v''} \ .
\ee

This argument can actually be easily generalized to arbitrary
shape functions $r_1$, as defined in \Eqref{eq:generaldefofa}.
We first include the parameter $b$ in the regulator, 
to account for the possibility to
rescale $k$
\begin{equation}
	R_k(z)=b k^2 a\, r_1(y/b) .
	\label{eq:generaldefofab}
\end{equation}
Then we choose $b=1/a$ such that the regulator becomes
\begin{equation}
	R_k(z)=k^2\, r_1(a y) .
	\label{eq:generalrescaledR}
\end{equation}
Then the $a\to0$ limit of \Eqref{eq:generaldefofab},
results in the constant regulator.
\footnote{By comparing this with the original regulator
	in \Eqref{eq:generaldefofa} we see that 
	we have effectively cast the regulator
	as a function of $k_\text{eff}^2=a k^2$,
	rather then of $k$ itself,
	and then considered $k_\text{eff}$ as $a$ independent.}

An alternative way of arguing that the $a\to0$ limit reduces the LPA flow equation
for the regulator (\ref{litima}) to 
the constant regulator case (\ref{cs}) is
by performing an $a$-dependent rescaling
as in Sec.~\ref{sec:oscill}.
Namely, by redefining 
\begin{subequations}
	\begin{align}
\tilde{\phi}&= a^{(d-2)/4} \hat{\phi}\ ,\\
v(\tilde{\phi})&= a^{d/2}\hat{v}(\hat{\phi})+a\frac{1}{(d-2)(4\pi)^{d/2}\Gamma(1+d/2)} \ ,
	\end{align}
\label{eq:adepresc}
\end{subequations}
in the flow equations for the regulator 
(\ref{litima}) and then taking the $a\to0$
limit at  fixed $\hat{\phi}$ and $\hat{v}$,
we again find \Eqref{cs}.
For instance in $d=3$ this rescaling entails
that the prefactor $a$ in (\ref{eq:flowPLAa3d})
goes away.

Both kind of arguments however are applicable 
only for non-exceptional $d$. In particular,
in some cases removing the momentum
	dependence of the regulator by sending $a\to0$,
	as in \Eqref{eq:rescaledlitim} and \Eqref{eq:generalrescaledR},
	is not possible, because
	the $a\to 0$ limit and 
	the momentum integral
	cannot be exchanged.
	This happens whenever the integral
	corresponding to the constant regulator
	is divergent.
	In fact,
the momentum integral leading to \Eqref{cs}
 is convergent only for $d<2$.~\footnote{
 However the UV divergence in $2\leq d<4$
 affects only the field-independent 
 part of the effective potential
 and in these cases it could be removed by implementing
 the standard subtraction as in 
 Sec.~\ref{sec:oscill}, see e.g.~\Eqref{eq:subtractedVflow}.
 Notice that this subtraction would introduce an IR divergence in $d=2$.
For some values of $d$ the limit
	$a\to0$ cannot be taken at the level
	of the integrands. In these cases we have first of all to 
	compute the integrals, and this requires to specify
	the shape function $r_1$.
	This is the main reason why in the present paper we 
	focus on the special regulator choice of \Eqref{litima}.
	More general results holding for arbitrary
	shape functions can be deduced once the 
	field-theory model, the number of Euclidean dimensions $d$ and the truncation of the EAA is specified.}

 If in the scalar LPA
we adopt the constant regulator in $d\geq2$,
 using analytic continuation 
 as a tool for the definition of the momentum integral
the result has a meromorphic structure with poles for even values of $d$. 
On the other hand, if we try to directly take the limit $a\to 0$
with the regulator (\ref{litima}), and expand the $Q$ functionals
(\ref{Qhyp}), with $n=d/2$, $d$ even and $\tilde m=0$,
in $a$ around $a=0$, there appear terms with $\log a$.
As a consequence, we expect that the
vanishing regulator limit of the LPA flow
equation will enjoy special properties
in even dimensions.
As a matter of fact, if analytic continuation
	is not adopted in the definition of the loop integrals,
	the arguments we just outlined point to the conclusion
	that the vanishing-regulator limit
	does not need to reproduce the constant-regulator
	case in the whole range $d\geq4$.

\subsection{Beta functions in two and four dimensions}
\label{sec:evend}

As we argued at the end of the previous section,
in the case of even dimensions the limit of
vanishing regulators has a more intricate structure.
Therefore, in this section we analyze
these special cases in more detail.

We start with $d=2$, where the flow equation of the LPA reads
\be
\partial_t v = -2 v + \frac{a}{4\pi(1-a)} \log \!\left( \frac{1+v''}{a+v''} \right).
\ee
Defining
	\begin{equation}
		v(\tilde{\phi}) = a \hat{v}(\tilde{\phi})
	\end{equation}
and simplifying a factor $a$ from the flow equation,
in the $a\to 0$ limit we are left with
\begin{align}
	\partial_t \hat{v} = -2 \hat{v} - \frac{1}{4\pi} \log a - \frac{1}{4\pi} \log \!\left( 1+\hat{v}'' \right).
\end{align}
The potential must be shifted by a factor that contains $\log a$,
i.e. $\hat{v} \to  \hat{v} - \frac{1}{8\pi} \log a$ ,
in order to eliminate this divergent term for the limit $a \to 0$ .
We observe that the coefficient of the 
$\log a$ term matches exactly the coefficient
of the $1/\epsilon$ pole of the expansion
of (\ref{cs})
 for $d=2+\epsilon$.~\footnote{This
 	correspondence 
 between $\log a$
 singularities of
 the flow equations for
 the regulator
 (\ref{litima})
 and $1/\epsilon$
 poles of
 (\ref{cs})
holds also in higher  even dimensions.}   
The finite logarithmic contribution coincides with the
one in \Eqref{cs}.
Therefore, up to a field-independent
shift of the potential, in $d=2$ the vanishing-regulator
limit agrees with the constant regulator.
\\

We then turn to the LPA in $d=4$.
We first truncate the potential to a polynomial
expansion around vanishing fields as
in \Eqref{potexp}.
For continuity with the previous sections, we also 
turn to the dimensionless 
couplings defined in \Eqref{eq:dlesslambdas}.
By considering the leading contributions 
to the beta functions $\tilde{\beta}_{2n}$
for vanishing $a$, 
we construct an ansatz based on the following scaling  
\begin{subequations}
	\begin{align}
		\hat{\lambda}_2&=a^{-1}\tilde{\lambda}_2\ ,\\
		\hat{\lambda}_4&=\log(a) \tilde{\lambda}_4\ ,\\
		\hat{\lambda}_{2n}&=a^{n-2}(\log a)^{n} \tilde{\lambda}_{2n},\quad n>2\ .
		\label{eq:hatlambda2n}
	\end{align}
	\label{eq:4Drescpol1}
\end{subequations}
Assuming the $\hat{\lambda}_{2n}$  couplings can be kept fixed in the $a\to 0$ limit,
results in the following set of beta functions
\begin{subequations}
	\begin{align}
	&	\partial_t\hat{\lambda}_2= -2\hat{\lambda}_2+\frac{\hat{\lambda}_4}{16\pi^2} \left[1+\frac{1+\log\!\left(
			1+\hat{\lambda}_2\right)}{\log a}
		\right],
			\label{eq:4Dpolybeta2}
		\\
	&	\partial_t\hat{\lambda}_4 =\frac{1}{\log a}\left[\frac{3}{16\pi^2}\frac{\hat{\lambda}_4^2}{1+\hat{\lambda}_2}+\frac{1}{16\pi^2}\hat{\lambda}_6\right],
			\label{eq:4Dpolybeta4}
		\\
	&	\partial_t\hat{\lambda}_6 =2\hat{\lambda}_6-\frac{15}{16\pi^2}\frac{\hat{\lambda}_4^3}{(1+\hat{\lambda}_2)^2}+\frac{1}{16\pi^2}\hat{\lambda}_8
		\nonumber\\
	&+\frac{\hat{\lambda}_8}{16\pi^2}
		\frac{1+\log\!\left(
			1+\hat{\lambda}_2\right)}{\log a}
		+\frac{15}{16\pi^2}\frac{\hat{\lambda}_4
		\hat{\lambda}_6}{(1+\hat{\lambda}_2)\log a}\ ,
			\label{eq:4Dpolybeta6}
	\end{align}
	\label{eq:4Dpolybeta}
\end{subequations}
and similar results for higher couplings.
Notice that terms of order $(\log a)^{-1}$
could be neglected as sub-leading in all beta functions apart for the second one, where 
such term is in fact the leading one.

In order to include the beta functions of all couplings in a  functional treatment, we turn to the task of including the definitions (\ref{eq:4Drescpol1}) in a rescaling of the effective potential.
It is impossible to achieve this goal by a two-parameters rescaling of the kind studied in the previous sections.
However, \Eqref{eq:hatlambda2n} trivially lends itself 
to a functional rescaling. Hence, we can treat the first
two couplings on a special footing, and embed the remaining
ones in a functional which is related to higher derivatives
of $v(\phi)$.

First, to simplify notations, it is convenient to define
\begin{align}
	\tilde{\rho}&=\tilde{\phi}^2/2\ ,\\
	u(\tilde{\rho})&=v(\tilde{\phi})\ .
\end{align}
Next, we define
\begin{equation}
	f(\tilde{\rho})=u'(\tilde{\rho})-\tilde{\lambda}_2
	-\frac{\tilde{\lambda}_4}{3}\tilde{\rho}\ .
	\label{eq:4Ddeff}
\end{equation}
So by construction $f(0)=f'(0)=0$, while $f^{(n)}(0)\propto\lambda_{2(n+1)}$.
The functional flow equation for $f$ can be obtained from the
functional equation for $u'$ by
\begin{equation}
	\partial_t f(\tilde{\rho})=\partial_t u'(\tilde{\rho})-\tilde{\beta}_{2}
	-\frac{\tilde{\beta}_{4}}{3}\tilde{\rho}\ .
\end{equation}
and then replacing $u_k'$ through the definition (\ref{eq:4Ddeff}).
The identities $\partial_t f_k(0)=\partial_t f_k'(0)=0$ also follow from this definition.
By the rescaling
\begin{align}
	f(\tilde{\rho})&=\frac{a}{\log a}\hat{f}(\hat{\rho})\ ,\\
	\tilde{\rho}&=a \log a\,\hat{\rho}\ ,
\end{align}
together with the previous definitions of $\hat{\lambda}_2$ and $\hat{\lambda}_4$,
we recover the full tower of relations (\ref{eq:4Drescpol1}).
By inserting the previous definitions in the flow equation for $u'(\rho)$
one can deduce the following functional flow equation
\begin{widetext}
\begin{align}
	\partial_t \hat{f}(\hat{\rho})&=-2\hat{f}(\hat{\rho})+2\hat{\rho}\hat{f}'(\hat{\rho})
	+\frac{3}{16\pi^2}\hat{f}'(\hat{\rho})
	+\frac{1}{8\pi^2}\hat{\rho}\hat{f}''(\hat{\rho})
	+\frac{5}{16\pi^2}\hat{\rho}\hat{f}''(0)
	-\frac{1}{16\pi^2}\frac{\hat{\rho}\hat{\lambda}_4^2}{1+\hat{\lambda}_2}
	\nonumber\\
	&-\frac{1}{16\pi^2}\hat{\lambda}_4	\log\!\left( 1+\hat{\lambda}_2\right)
	+\frac{1}{16\pi^2}\hat{\lambda}_4\log\!\left( 1+\hat{\lambda}_2+\hat{\lambda}_4 \hat{\rho}\right).
	\label{eq:betaf}
\end{align}
\end{widetext}
This functional flow generates the leading terms in  Eq.~(\ref{eq:4Dpolybeta6}),
and similar beta functions for the higher-order
couplings,
upon truncating it to a polynomial ansatz regular at the origin.
However, we stress again that \Eqref{eq:betaf}
does not include Eqs.~(\ref{eq:4Dpolybeta2},\ref{eq:4Dpolybeta4}),
which therefore have to be supplemented to exhaust the LPA
flow equations.

These flow equations are different from those of a constant regulator.
Indeed, the latter are formally UV divergent.
More specifically, in $\tilde \beta_{2n}$ the contribution
linear in $\tilde\lambda_{2n+2}$ corresponds to a momentum
integral with dimension $2$
which is not regularized by the constant regulator~\eqref{eq:CSreg}.
Similar discrepancies arise in $d=6,8,\dots$.
The flow equation for the constant regulator in $d=4-\epsilon$ reads
\begin{align}
	\partial_t { v}=&\ -4 { v}+2 \tilde{\rho}  {v}'
	\nonumber\\
	&+\frac{\left(2 \tilde{\rho}  {v}''+{v}'+1\right) \left[\log\! \left(2 \tilde{\rho}  {v}''+{v}'+1\right) -1\right]}{16 \pi ^2}
	\nonumber\\
	&
	+\frac{\left(2 \tilde{\rho}  {v}''+{v}'+1\right)}{16 \pi ^2}
	\left[\gamma -\log (4\pi )-\frac{2}{\epsilon}\right].
\label{eq:csd4}
\end{align}
The third line in this equation
arises from the expansion of the sine in
the denominator of \Eqref{cs}.
It provides contributions to the ${\tilde \lambda}_{2n+2}$
term inside ${\tilde \beta}_{2n}$.
Such terms would be absent in the \MSbar scheme.
These $1/\epsilon$ contributions which are divergent in $d=4$ 
	are a typical product of the analytic continuation
	adopted in the definition of the integral.
	Similar contributions which diverge in $d=4$ are expected
	also if any other alternative definition is chosen.
    For instance, if a sharp UV cutoff $\Lambda$ is introduced,
    the third line of \Eqref{eq:csd4} would be replaced by a
    different expression which is ill-defined in the 
    $\Lambda\to\infty$ limit.

If we perform an \textit{ad hoc} subtraction
of the third line, the flow equation
\eqref{eq:csd4} leads to  the following 
beta functions
\begin{subequations}
\begin{align}
	{\tilde \beta}_2=&\ -2 {\tilde \lambda}_2
	+ \frac{{\tilde \lambda}_4 \log\! \left({\tilde \lambda}_2+1\right)}{16 \pi ^2}\ ,
\end{align}
\begin{align}
	{\tilde \beta}_4=&\ \frac{3 {\tilde \lambda}_4^2}{16 \pi ^2 \left({\tilde \lambda}_2+1\right)}+\frac{{\tilde \lambda}_6 \log\! \left({\tilde \lambda}_2+1\right)}{16 \pi ^2}\ ,\\
{\tilde \beta}_6=&\ 2 {\tilde \lambda}_6 -\frac{15 {\tilde \lambda}_4^3}{16 \pi ^2 \left({\tilde \lambda}_2+1\right)^2}
+\frac{{\tilde \lambda}_8 \log\! \left({\tilde \lambda}_2+1\right)}{16 \pi ^2}
\nonumber\\
&+\frac{15 {\tilde \lambda}_6 {\tilde \lambda}_4}{16 \pi^2 \left({\tilde \lambda}_2+1\right)}\ .
\end{align}
\label{eqs:4Dcspolybeta}
\end{subequations}
A comparison with \Eqref{eq:4Dpolybeta} 
immediately reveals several differences.
Apart for the scaling (classical) terms,
the first two quantum/statistical terms are equal, 
up to the fact that the $\lambda_2$ dependence
of the $\lambda_{2n+2}$ term has been
washed away in \Eqref{eq:4Dpolybeta} by the $a\to 0$ limit,
and up to the crucial $\log a$
dependence of \Eqref{eq:4Dpolybeta4}.
However, all the additional quantum/statistical terms in
\Eqref{eqs:4Dcspolybeta} are absent
in \Eqref{eq:4Dpolybeta}.

The peculiar simplicity which the Eqs.~\eqref{eq:4Dpolybeta}
attain in the $a\to0$ limit,
together with the $1/\log a$ dependence
of \Eqref{eq:4Dpolybeta4}, raises the 
question as to whether these
beta functions retain enough
physical information for being practically
useful. As a first step towards
addressing this question,
we limit ourselves to a simple observation.
Namely, as long as the subleading
logarithmic  $a$-dependence
is retained in \Eqref{eq:4Dpolybeta},
the $\phi^4$-theory beta function
and other universal physics
is still present.
For instance, we can
 study the WF
fixed point in $d=4-\epsilon$.
In order to employ \Eqref{eq:4Dpolybeta}
in this study, 
we need to prescribe that 
the $\epsilon\to 0$
limit be taken before the $a\to 0$ one.
This means in practice that
the vanishing-regulator limit is taken
on the $d=4$ FRG equations.
Had we sent $a\to 0$ in $d<4$, we would have found different
equations for $\hat{\lambda}_{2n}$
and precisely the constant-regulator ones,
as already mentioned in Sec.~\ref{sec:vanishingVScs}.

Within the simplest truncation corresponding
to retaining only $\tilde{\lambda}_2$ and
$\tilde{\lambda}_4$, where we add the 
classical scaling term $-\epsilon\lambda_4$
to $\hat{\beta}_4$ to account for
the shift of dimensionality, 
the WF fixed point
to first order in $\epsilon$ is located at
\begin{align}
	\hat{\lambda}_2=&\ \frac{1}{6}\epsilon\left(1+\log a\right)\ ,\\
	\hat{\lambda}_4=&\ \frac{16}{3}\pi^2\epsilon\log a\ .
\end{align}
These fixed-point couplings 
have to be interpreted as small,
even if they seemingly blow up 
for $a\to0$, because the
limit $\epsilon\to 0$ has to be taken first.
Notice that keeping
the sub-leading order-$(\log a)^{-1}$ 
contribution to $\hat{\beta}_4$
is essential for revealing
the fixed point.
By computing the corresponding critical
exponents, we find the 
universal one-loop result
\begin{equation}
\theta_1=2-\frac{\epsilon}{3}\ ,\qquad
\theta_2=-\epsilon \ .
\end{equation}

\section{\texorpdfstring{$\boldsymbol{O( N\! +\! 1)}$}{O(N+1)} symmetry in nonlinear models}
\label{sec:ON+1}

As a first example of a symmetry that is broken by the regulator,
we shall consider here the two dimensional $O(N+1)$ nonlinear sigma model
in a particular coordinate system.
We start from the order-$\partial^2$ expansion of $\Gamma_k$
for a $O(N)$-invariant multiplet of scalars
\begin{align}
\Gamma_k[\phi] =&\  \int\! d^2x\, \Big[ U_k(\rho) 
+ \frac{1}{2}Z_k(\rho) \partial_\mu \phi_a \partial^\mu \phi^a
\nonumber\\
& + 
\frac{1}{4}Y_k(\rho) \partial_\mu \rho \partial^\mu \rho \Big]\ ,
\label{eq:GammaDE}
\end{align}
where the $N$ fields $\phi^a$ are in
the fundamental representation of $O(N)$,
and $\rho=\phi^a\phi^a/2$ is the corresponding local invariant.
We further define the radial wave function renormalization
\be
\tilde Z_k \left( \rho \right) = Z_k\left( \rho \right)+\rho Y_k\left( \rho \right),
\label{eq:tildeZ}
\ee
The beta functions for $Z_k$, $\tilde Z_k$ and $U_k$
are given in App.~\ref{app:floweqDE}.

If we make the assumptions
\begin{subequations}
\begin{align}
Z_k(\rho)&=\frac{Z_k}{g_k^2}\ ,\\
\tilde Z_k (\rho)&=\frac{1}{g_k^2} \left( \frac{1}{Z_k}-2 \rho \right)^{-1}\ ,\\
U_k&=0\ ,
\end{align}
\label{eq:nlsmZV}
\end{subequations}
the EAA becomes 
\be
\Gamma_k[\phi] =  \int d^dx 
\frac{Z_k}{2g_k^2}\left(\delta_{ab}
+\frac{\phi_a\phi_b}{\frac{1}{Z_k}-2\rho}
\right)
\partial_\mu \phi^a \partial^\mu \phi^b
\ .
\label{eq:GammaNLSM}
\ee
The tensor in parentheses is the metric of the $N$-dimensional sphere
of radius $Z_k^{-1/2}$,
written in a coordinate system that consists in projecting a point
of the sphere orthogonally on the equatorial plane.
In this way the northern hemisphere is mapped to the
domain $\phi^a\phi^a<1/Z_k$.
The symmetry group is extended to $O(N+1)$.

A standard cutoff
\be
\Delta S_k(\phi) =  \frac{Z_k}{2g_k^2}\int d^2x \,
\phi_a R_k(-\partial^2) \phi^a
\nonumber
\ee
breaks $O(N+1)$ invariance, while preserving $O(N)$.
Therefore, if we start at some scale $k$ with an EAA
of the form (\ref{eq:GammaNLSM}),
the flow will immediately generate $O(N+1)$-violating terms,
and thus will take place in the larger theory space
parameterized by (\ref{eq:GammaDE}).

This can be seen already
by projecting the flow generated by the ansatz 
(\ref{eq:GammaNLSM}) on the local potential,
i.e.~by considering \Eqref{eq:DEbetaUQ}.
For nonvanishing $a$ and for field-dependent wave function
renormalizations, the choice $U_k=0$
is not preserved by the RG flow.
However, in the $a\to0$ limit it  indeed becomes a
consistent ansatz,
as in $\partial_t U_k$ the rhs~behaves like $a\log a$
when $a\to 0$.

Let us then inspect the flow of the wave function
renormalizations. Inserting the previous ansatz
in the flow equation (\ref{eq:DEbetaZQ})
for $\tilde{Z}_k(\rho)$,~\footnote{Note 
that now $Z_k(\rho =0) = Z_k/g_k^2$, so inside the formulae for the $Q$ functionals 
we must send $Z_k \to Z_k/g_k^2$ and $\eta_k \to \eta_k + 2 \partial_t g_k/g_k$.} in the limit $a \to 0$ we obtain
\begin{align}
&-\frac{2 Z_k \partial_t g_k }{g_k^3 \left( 1 -2 Z_k \rho \right)}-\frac{Z_k \eta_k }{g_k^2 \left(1-2 Z_k \rho \right)^2}
= \frac{Z_k}{4\pi}
\nonumber\\
&\times\frac{ (2 \partial_t g_k +(\eta_k -2) g_k) (2 (N-1) Z_k \rho +1)}{g_k (1-2 Z_k \rho )^2}
.
\end{align}
As the functional $\rho$ dependence 
on both sides of the equation is comparable,
this equation can be algebraically solved 
for $\partial_t g_k$ and $\eta_k$,
resulting in
\begin{subequations}
 \begin{align}
\partial_t g_k &= -\frac{(N-1) g_k^3}{4\pi+g_k^2} \ ,
\\
\eta_k &= -\partial_t \log Z_k = \frac{2 N g_k^2}{4\pi + g_k^2} 
\ .
\end{align}
\end{subequations}
These are the correct one-loop beta functions,
augmented by RG resummations due to the
dependence of the regulator on $Z_k$ and $g_k$. 
The same result can be derived by considering
the flow equation for $Z_k(\rho)$.
Thus, within the present truncation
the nonlinearly realized
$O(N+1)/O(N)$ symmetry is preserved by
taking the limit $a \to 0$.
Essentially the same flow equations have
been obtained in Ref.~\cite{BPZ}
with a pseudo-regulator reproducing the \MSbar scheme.~\footnote{The
	 present result 
	is obtained by setting $\sigma=1$
in the beta functions of Ref.~\cite{BPZ}.}\\

The assumption $U_k=0$, although justified by
the observation that only a trivial potential 
is compatible with the nonlinearly realized 
symmetry, can be easily relaxed as long as
this explicit symmetry-breaking term
is treated as an external source.
The simplest
of such terms is a linear coupling to the
$O(N+1)/O(N)$ variation of $\phi^a$,
i.e.~$\phi^{N+1}$
\begin{equation}
 U_k =-H \sqrt{\frac{1}{Z_k}-2\rho}\ .
 \label{eq:Uofh}
\end{equation}
This ansatz, comprehending an arbitrary source $H$,
was observed to be compatible
with the flow equation in the case of an \MSbar
pseudo regulator \cite{BPZ}.
This linear term can also be used to construct
an exact FRG equation which manifestly preserves
the full $O(N+1)$ symmetry for every regulator
function $R_k$, see App.~\ref{app:masterON+1}.
For the present standard FRG implementation and 
regularization scheme, the ansatz (\ref{eq:Uofh})
is not compatible with the flow equation of
the potential, neither for $a\neq0$
nor in the $a\to 0$ limit.
Only by assuming that $H$ be a function of $a$ 
vanshing faster than $a$ itself, closure of
the $O(\partial^2)$ RG flow on the ansatz (\ref{eq:GammaNLSM})
is recovered.
To understand this phenomenon it is
necessary to study how the modified master
equation for the $O(N+1)/O(N)$ symmetry
behaves in the $a\to0$ limit, which is the topic
of Appendix \ref{app:masterON+1}.
There we show how the construction of a nonvanishing
potential term for the nonlinear sigma model
is indeed a complicated problem which 
requires the simultaneous solution
of both the flow equation and the 
modified master equation. As
explained in Appendix \ref{app:masterON+1},
in solving this problem the $a\to0$ limit is of limited use.

\section{Background field issues}
\label{sec:background}

When one splits the field into a classical background 
and a  quantum/statistical fluctuation
\be
\phi=\phi_B+\varphi\ ,
\label{split}
\ee
the action, being a function of $\phi$, is invariant
under the shift symmetry
\begin{subequations}
\begin{align}
&\phi_B \mapsto \phi_B+\epsilon \ ,
\\
&\varphi \mapsto \varphi-\epsilon\ .    
\end{align}
\label{shift}
\end{subequations}
This can be expressed by the identity
\be
\frac{\delta S}{\delta \phi_B}-\frac{\delta S}{\delta \varphi}=
0\ .
\label{clward}
\ee
On the other hand, the regulator only depends on the 
background field
and is therefore not invariant under the split symmetry.
In particular in gauge theories, in order to preserve background gauge invariance, the cutoff is usually written as a function of the background covariant derivative: $R_k(-\bar D^2)$.
This effect can be mimicked in the scalar case by artificially introducing a dependence of $R_k$ on $\phi_B$.
For example, Morris and collaborators considered
regulators of the general form \cite{Bridle:2013sra}
\be
R_k(z) = (k^2-k^2 h(\tilde\phi_B)-z) \theta(k^2-k^2 h(\tilde\phi_B)-z) \ .
\label{regwithhA}
\ee
The EA then becomes a functional 
$\Gamma_k[\varphi,\phi_B]$,
i.e.~it has a separate dependence on these two arguments.
The breaking of the shift symmetry results in a modified Ward identity
\be
\frac{\delta \Gamma_k}{\delta\phi_B}
-\frac{\delta \Gamma_k}{\delta\varphi} = 
\frac{1}{2}\Tr\left[\left(\frac{\delta^2 \Gamma_k}{\delta\varphi \delta\varphi}+R_k\right)^{-1}\frac{\delta R_k}{\delta\phi_B}\right]\ .
\label{ward}
\ee
It has been shown that such background dependence in the regulator
can either destroy physical fixed points or create artificial ones~\cite{Bridle:2013sra}.
On the other hand, when the FRG equation (\ref{FRGE}) is solved together with
the Ward identity (\ref{ward}),
the correct physical picture can be reconstructed.
While this can be achieved in the scalar case~\cite{Bridle:2013sra},
it is much harder in the case of gauge theories,
and in particular for gravity~\cite{Dietz:2013sba}.
It is therefore desirable to find other ways around this obstacle.
The form of the equation (\ref{ward}) 
suggests that
in the limit of vanishing regulator the shift symmetry is restored.
One would therefore expect that in this limit 
the aforementioned pathologies should also disappear.
In this section we will see
how this actually happens in the scalar theory.

We begin by briefly reviewing some results of Ref.~\cite{Bridle:2013sra}.
We consider the same system as in Sec.~\ref{WFLPAvana},
in $d=3$,
but we use the regulator (\ref{regwithhA}).
In a single-field approximation one
identifies $\phi_B=\varphi$. The
corresponding flow equation for the potential reads
\begin{align}
&\partial_t v =- 3 v +\frac{1}{2}\tilde\varphi v' 
\nonumber\\
&+ \frac{\left( 1- h \right)^{3/2}}{1- h + v''}  \left(\!1\!-\!h-\!\frac{1}{2}\partial_t h\!+\!\frac{1}{4} \tilde\varphi h' \! \right) \theta(1-h) \,.
\label{o13dwithh}
\end{align}
Two special cases for $h$ have been considered.
The first case is $h=\alpha\tilde\varphi^2$.
In this case, for $\alpha<0$ the Heaviside theta on the rhs of \Eqref{o13dwithh} is equal to one.
Solving the fixed point equation, one finds that
the Gaussian fixed point becomes interacting
and an increasing number of fake fixed points appear,
as $\alpha$ becomes more negative.
For example, Tab.~\ref{fig:spurious} presents 
the nontrivial fixed points and
the associated relevant critical exponents
for two negative values of $\alpha$.
\begin{table}
		\centering
\begin{minipage}{.3\columnwidth}
	\centering
	\begin{tabular}{ |c|c|c| } 
		\hline
		FP & $\theta_1$ & $\theta_2$  \\ 
		\hline
		$1$ & $1.17$ & -  \\ 
		$2$ & $2.11$ & $0.82$  \\ 
		\hline
	\end{tabular}\\
   $\boldsymbol{\alpha=-1/2}$
\end{minipage}
\ \,
\begin{minipage}{.48\columnwidth}
	\centering
\begin{tabular}{ |c|c|c|c|c| } 
	\hline
	FP & $\theta_1$ & $\theta_2$ & $\theta_3$ & $\theta_4$ \\ 
	\hline
	$1$ & $0.89$ & - & - & - \\ 
	$2$ & $2.35$ & $0.76$ & - & - \\ 
	$3$ & $2.02$ & $1.43$ & $0.60$ & - \\ 
	$4$ & $2.10$ & $1.69$ & $1.08$ & $0.39$ \\ 
	\hline
\end{tabular}\\
   $\boldsymbol{\alpha=-2}$
\end{minipage}
\caption{The nontrivial fixed-point solutions
of \Eqref{o13dwithh} with $h=\alpha\tilde{\phi}^2$, and the corresponding relevant
critical exponents, for ${\alpha=-1/2}$
(left panel) and ${\alpha=-2}$ (right panel). The 
entries which are
left blank correspond 
to irrelevant deformations.
FP${}_1$ is
the Wilson-Fisher fixed point,
while FP${}_2$
is a "deformed Gaussian" fixed
point as it possesses 
two relevant directions.}
\label{fig:spurious}
\end{table}
In both cases FP${}_2$ is the deformed Gaussian fixed point.
For $\alpha>0$ because of the Heaviside theta function
on the rhs of \Eqref{o13dwithh}, 
$v = A \tilde\varphi^6$ for $\tilde\varphi>1/\sqrt{\alpha}$.
The Gaussian fixed point is always absent,
and for $\alpha > 0.08$ also the WF fixed point disappears.

The second case is $h=\alpha  v''$.
The Gaussian~\footnote{Note that the Gaussian fixed point corresponds to the point $(v(0),v'(0))=(1/3,0)$ .} and the WF fixed points always exist, but when $\alpha$ is increased,
new fixed points appear near the Gaussian one~\footnote{In particular for $\alpha \ge 0.85$ a first additional fixed point appears. } 
and move away from it as $\alpha$ becomes bigger: for example, for $\alpha=1$ there is a spurious fixed point and for $\alpha=2$ there are three of them.

In Ref.~\cite{Bridle:2013sra}
the authors  solve
the anomalous Ward identity for shift symmetry
and show how to recover the physical results.
Instead, we shall discuss here the effect of
taking the limit of vanishing regulator.
To this end, we first introduce the parameter $a$ in
(\ref{regwithhA})
\be
R_k(z) = a (k^2-k^2h(\tilde\phi_B)-z) \theta(k^2-k^2h(\tilde\phi_B)-z) \ .
\label{regwithah}
\ee 
Within a single-field LPA truncation this leads to the flow equation
\begin{widetext}
\begin{align}
\partial_t v = - 3 v +\frac{1}{2}\tilde\varphi v' + 
\theta(1-h)\frac{a\left( 1- h \right)^{3/2}}{a(1-h)+ v''}
 \left(1-h-\frac{1}{2}\partial_t h+\frac{1}{4} \tilde\varphi h'  \right)
 {}_2F_{1}\! \left(1,\frac{3}{2},\frac{5}{2}; \frac{(a-1)(1-h)}{a(1-h)+ v''}\right) \,.
\label{o13dwithah}
\end{align}
\end{widetext}
Again, we discuss separately the two choices for the function $h$.

\begin{figure}[H]
	\centering
	\includegraphics[width=.97\columnwidth]{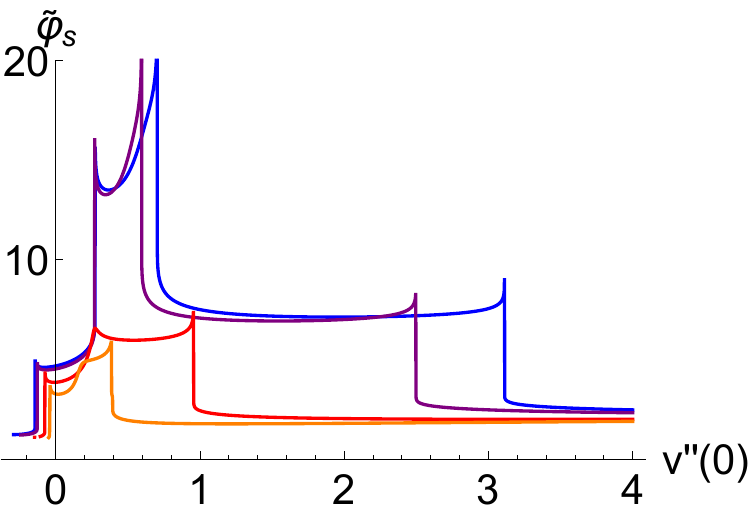}
	\caption{Spike plots with $h=-2\tilde{\varphi}^2$ and with different values of $a$:  $a=0.5$ (blue), $a=0.4$ (purple), $a=0.16$ (red) and $a=0.07$ (orange).
		For each curve, the rightmost spike is the deformed Gaussian fixed point, the leftmost one is the Wilson-Fisher fixed point.
		Decreasing $a$ further, both these fixed points move towards the origin.
		In the red and in the orange curves one and two of the fake fixed points have disappeared correspondingly.}
	\label{fig:spikes}
\end{figure}

\paragraph{\textbf{First case:} $\boldsymbol{h=\alpha\tilde\varphi^2}$.}
Following Ref.~\cite{Bridle:2013sra}
we start with a quadratically-field-dependent regulator.
However we slightly depart from
that reference in that we
find it more convenient to portrait
the landscape of fixed points
by a different numerical method,
a shooting from the origin.
This consists in constructing
numerical  solutions for
each possible value of the
boundary condition $v''(0)$.
The generic solutions however
terminate at a finite value $\tilde\varphi_S$
of the field  which corresponds
to a movable singularity of the fixed-point
equation.
In this process one obtains a plot of $\tilde\varphi_S$ as a function
of $v''(0)$ (also known as spike plot).
Sharp maxima in the latter variable are in one-to-one 
correspondence with
global fixed points. 
The result is presented in Fig.~\ref{fig:spikes}.
We can see that decreasing $a$ the spurious fixed points disappear
and the physical fixed points converge to the origin.
This is the same phenomenon that we observed in Secs.~\ref{WFLPArel} and \ref{WFLPAvana}.

At these fixed points,
we compute the spectrum of 
critical exponents 
with the same method used in
Ref.~\cite{Bridle:2013sra}, 
namely by shooting
from infinity,
as we did in Sec.~\ref{WFLPAvana}.
This means that we
first construct
an asymptotic expansion
of the fixed-point
potential as well as of the
eigenfunction of the
linearized flow around the
fixed point.
For $\alpha<0$ the Heaviside theta on the rhs of equation (\ref{o13dwithah}) is equal to one, and the potential has the following behavior at infinity
\begin{widetext}
\begin{align}
v =&\ A \tilde\varphi ^6+ \frac{a |\alpha|^{5/2}}{150 A}|\tilde\varphi| +\frac{a|\alpha|^{3/2}(525A-(3+2a)\alpha^2)}{31500 A^2 |\tilde\varphi|}
\nonumber\\
& +\frac{a \sqrt{-\alpha } \left(212625 A^2-\alpha ^2 (3780 a A+5670 A)+\left(16 a^2+24 a+30\right) \alpha ^4\right)}{17010000 A^3 |\tilde\varphi|^3}+ O\left( |\tilde\varphi|^{-5} \right) \,.
\end{align}
\end{widetext}
Shooting on $A$ and on a
corresponding asymptotic
parameter for the perturbation,
and by demanding
$\mathbb{Z}_2$ parity at
the origin, we determine
the location of the 
fixed point as well as
the quantized values of
the critical exponents.
In the $a\to 0$ limit the latter become independent of 
$\alpha$ and agree with
the spectrum discussed
in Sec.~\ref{WFLPAvana}.

For $\alpha>0$, because of the Heaviside theta one the rhs of \Eqref{o13dwithah} $v = A \tilde\varphi^6$ for $\tilde\varphi>1/\sqrt{\alpha}$.
Therefore for $\tilde\varphi<1/\sqrt{\alpha}$ the potential 
 as a function of 
$\delta=\left( \frac{1}{\sqrt{\alpha}}-\tilde\varphi \right)^{1/2}$
 has the following asymptotic behavior
\begin{align}
v=&\ \frac{A}{\alpha^3}-\frac{6A}{\alpha^{5/2}} \delta^2+ \frac{15 A}{ \alpha^2} \delta^4+ \frac{2\sqrt{2} \,a\,\alpha^{13/4}}{75A}\delta^5
\nonumber\\
& -\frac{ 135000A^4+a^2\alpha^{10}}{6750 \alpha^{3/2}A^3} \delta^6+o\left( \delta^7 \right) \ .
\end{align}
Shooting from infinity and decreasing $a$ we recover the Gaussian and the WF fixed points.
In particular, for $\alpha =1/25$ the Gaussian fixed point reappears for $a \lesssim 10^{-2}$, while for $\alpha =1/9$ the WF fixed point reappears for $a\lesssim 0.35$ and the Gaussian one for 
$a \lesssim 4 \cdot 10^{-3}$.
Also in this case the critical exponents of the Gaussian and WF fixed points approach the
values found for vanishing $a$
in Sec.~\ref{WFLPAvana}.

\paragraph{\textbf{Second case:} $\boldsymbol{h=\alpha v''}$.}
We then move on to consider
a regulator which depends on
the second derivative of the effective potential, through a constant $\alpha>0$.
In this particular case
shooting from the origin is
not convenient for technical
reasons, therefore we 
shoot from large field values. 

This time  $v = A \tilde\varphi^6$
for $\tilde\varphi>\tilde\varphi_c\equiv(30A \alpha)^{-1/4}$
provided $v''>1/\sqrt{\alpha}$. 
Below $\tilde\varphi_c$ the potential can be expanded
in $\delta=\tilde\varphi_c-\tilde\varphi$ as follows
\begin{align}
v =&\ \frac{A}{\left( 30A\alpha \right)^{3/2}}
- \frac{6A}{\left( 30A\alpha \right)^{5/4}}  \delta + \frac{1}{2\alpha^2}\delta^2+ F(\delta) \,,
\\
F =&\ \delta ^{16/5} \Bigg( -\frac{25 \sqrt{5} A^{1/10} \alpha^{-17/10}}{88 \sqrt{2}\, 3^{3/10}a^{2/5}} 
\nonumber\\
&+
\frac{125\ 5^{3/4} A^{-1/20} \alpha^{-53/20}}{5984 \sqrt[4]{2}\, 3^{17/20} a^{4/5} } \delta^{1/5} \nonumber\\
&
-\frac{71875 A^{-1/5} \alpha^{-18/5}}{246445056\ 3^{2/5} a^{6/5}}\delta^{2/5} + o(\delta^{3/5})\Bigg) \,.
\end{align}

Shooting on $A$ and searching for
values which correspond to
a vanishing
$v'(0)$ one can reveal
several spurious fixed
points at nonvanishing
 $\alpha$ and $a$.
More and more of them are generated from the Gaussian fixed point for bigger and bigger 
values of $\alpha$.
We find that decreasing $a$
at fixed $\alpha>0$ reduces the
number of spurious fixed points,
and in the $a\to0$ limit all
of them disappear while the Gaussian and the WF fixed points merge.
We verify that also in this case the critical exponents tend to the
 values obtained
 in Sec.~\ref{WFLPAvana} for $a\to0$.

\subsection{Ward Identity for the shift symmetry}

Going beyond a single-field approximation,
i.e.~keeping both $\varphi$ and $\phi_B$
as distinct, 
the LPA truncation 
 becomes~\footnote{The mixing term $\partial_\mu \phi_B \partial^\mu \phi$ is ruled out by the $\mathbb{Z}_2\times \mathbb{Z}_2$
symmetry on the arguments of the EAA.}
\begin{align}
\Gamma_k [\varphi , \phi_B]
 =&\ \int\! d^d x\, \left( \frac{1}{2}\left( \partial_\mu \varphi \right)^2 
+ \frac{1}{2}\left( \partial_\mu \phi_B \right)^2 
\right. \nonumber\\ & \left. 
+ V_k(\varphi , \phi_B) \right) \ .
\end{align}
Using the regulator \eqref{regwithah}
the modified Ward identity (\ref{ward})
and the flow equation become
\begin{widetext}
\begin{align}
& \partial_{\tilde\varphi} v - \partial_{{\tilde\phi}_B} v =
c_d \frac{  h' }{2}\, \frac{a(1- h)^{d/2}}{a(1-  h) + \partial_{\tilde\varphi}^2 v} 
\ {}_2F_1\! \left( 1, \frac{d}{2} , \frac{d}{2}+1 ; -\frac{(1-a)(1- h))}{a(1-  h) +\partial_{\tilde\varphi}^2 v} \right) ,
\\
& \partial_t v +d \, v- 
\frac{(d-2)}{2} \left( {\tilde\varphi} \partial_{\tilde\varphi} v + {\tilde\phi}_B \partial_{{\tilde\phi}_B} v \right) 
=  \nonumber\\
&c_d \, \frac{a(1-h)^{d/2}}{a(1-  h) + \partial_{\tilde\varphi}^2 v} 
\left(1- h-\frac{1}{2}\partial_t  h + \frac{1}{4} \left( d-2 \right) {\tilde\varphi}  h'  \right)
\ {}_2F_1\! \left( 1, \frac{d}{2} , \frac{d}{2}+1 ; -\frac{(1-a)(1- h))}{a(1-  h) +\partial_{\tilde\varphi}^2 v} \right)
\!,
\end{align}
\end{widetext}
where 
$c_d = \left( (4\pi)^{d/2} \Gamma\left( \frac{d}{2}+1 \right) \right)^{-1}$ .
We rescale all the quantities in the following way
\begin{align}
{\tilde\varphi} &=  a^{(d-2)/4} \,\hat{\varphi} \,,
\\ 
{\tilde\phi}_B &=  a^{(d-2)/4} \,\hat{\phi}_B \,,
\\
 v(\tilde{\varphi}) &= a^{d/2} \, \hat{v}(\hat{\varphi}) + a \frac{1 }{(4\pi)^{d/2} (d-2)\,\Gamma\left( \frac{d}{2}+1 \right)} \,,
\\
h &= a^\gamma \, \hat h\ .
\end{align}
This set of definitions agrees with
the one in \Eqref{eq:adepresc}.
Here $\gamma$ depends on the choice of $h$: 
for example $\gamma = 1$
for both $h = \alpha {\tilde\phi}_B^2 $ and $ h = \alpha v''$.
For the sake of generality we shall keep $\gamma$ free
for the time being.
Expanding for small $a$ and assuming 
$2<d<4$, the Ward identity and the flow equation become
\begin{align}
&\partial_{\hat\varphi} {\hat v} - \partial_{{\hat\phi}_B} {\hat v} =
\frac{a^{\gamma+1-d/2}}{d(4\pi)^{d/2}\Gamma\left( \frac{d}{2} \right)} 
\,  \hat{h}'
+\cdots \,,
\\
& \partial_t {\hat v} +d \, {\hat v}- \frac{(d-2)}{2} \left( {\hat\varphi} \partial_{\hat\varphi} {\hat v} + {\hat\phi}_B \partial_{{\hat\phi}_B} {\hat v} \right)
=
\nonumber\\
&
-\frac{a^{\gamma+1-d/2} }{d(4\pi)^{d/2} \Gamma\left( \frac{d}{2} \right)} \, 
\left( \partial_t  \hat{h} +d \,  \hat{h} - \frac{(d-2)}{2} {\hat\phi}_B \, \hat{h}'  \right)
\nonumber\\  
&
+\frac{\Gamma\left( \frac{d}{2}-1 \right)}{(4\pi)^{d/2}}
\,(1 + \partial_{\hat\varphi}^2 {\hat v})^{d/2-1}  
+\cdots \,,
\end{align}
where the dots denote quantities that go to zero for $a \to 0$ .\\
From the modified Ward identity we see that 
to have a well defined 
vanishing-regulator limit
we must demand $\gamma \ge \frac{d}{2}-1$.
If $\gamma > \frac{d}{2}-1$ ,
$\partial_{\hat\varphi} {\hat v} = \partial_{{\hat\phi}_B} {\hat v}$ : this implies that 
${\hat v}({\hat\varphi}, {\hat\phi}_B) = {\hat v}({\hat\varphi}+{\hat\phi}_B)$ and so we recover the shift symmetry and the flow equation without background.
If $\gamma=\frac{d}{2}-1$,
the modified Ward identity gives
\be
{\hat v}({\hat\varphi} , {\hat\phi}_B) = {\hat v}_s({\hat\varphi} + {\hat\phi}_B)
- \frac{1}{d(4\pi)^{d/2}\Gamma\left( \frac{d}{2} \right)} 
\hat h({\hat\phi}_B) \,.
\ee
Inserting this result into the flow equation, we recover again 
the equation without background.

\section{Discussion}
\label{sec:discussion}

We have discussed the effect of an overall suppression of the
regulator with a constant factor $a$, and in particular
the limit $a\to0$, that we called the limit of vanishing regulator.
Let us summarize the main results.

As is clearly seen already in the case of the quantum mechanical
oscillator, decreasing $a$ has the effect of accelerating the flow,
in the sense that already a small decrease of $k$ leads
very fast to the effective action. Thereafter things remain
nearly constant with $k$.
However, the quantum-mechanical study 
shows that in general different results are  obtained
depending on whether the $a\to0$ limit is 
performed on the solutions of the flow equations,
or on the beta functions themselves, see Fig.~\ref{fig:trajectoriesa}.
While the former way of taking the
limit is rather straightforward, 
obtaining meaningful results from the
latter process requires a suitable
$a$-dependent rescaling of the couplings.

In the case of the Wilson-Fisher fixed point,
we have first studied the
first form of the vanishing-regulator limit,
by analysing the $a$ dependence of the 
fixed-point solution.
Decreasing $a$ has the effect of shifting the fixed points towards
the Gaussian one, but the scaling exponents remain distinct
even in the limit $a\to 0$.
Here we have limited our analysis to the leading order of
the derivative expansion.

In a polynomial approximation of the potential,
the values of the scaling exponents become progressively worse
as one increases the order of the polynomial.
This is in agreement with the statement in Ref.~\cite{Litim:2002cf}
that the radius of convergence of the Taylor expansion of $V$
is proportional to $a$.
We have avoided this problem by also considering the
functional treatment (LPA), but in this case one gets
the exponent $\nu=1$, which is worse than for any polynomial
and coincides with the upper boundary 
conjectured in \cite{Litim:2002cf}.

We have then analyzed the second form
of the vanishing-regulator limit,
taking it on the LPA beta functional of 
scalar field theory, finding
agreement with the first kind
of limit as far as the critical
exponents are concerned,
although the locations of the
fixed point differ.
Even though
some naive arguments suggest that 
the limit of vanishing regulator might generally
reproduce the results of a constant (momentum independent)
mass-like regulator, we have observed that in the LPA
this is the case 
 only when the constant-regulator 
	momentum integrals are convergent.
As we adopted analytic continuation in the definition 
of the integrals, this excludes even integer values of $d\geq4$
(the $d=2$ case can indeed be reduced to the constant-regulator
case by a field-independent shift in the potential).
As a consequence, the vanishing-regulator limit
remains different from the constant regulator in $d=4$.
We expect this conclusion to hold also in higher even dimensions,
if analytic continuation is used, or in
	the whole range $d\geq4$ without analytic continuation.
It remains to be seen whether
these conclusions are robust against
enlargements of the truncation.
For instance, at the second order
of the derivative expansion, there might be a nontrivial interplay
between the momentum-derivatives of the regulator  and the
$a\to0$ limit, resulting in further differences
between the constant and the vanishing regulators.

For all these reasons, it will be quite interesting to systematically study the next order of the derivative expansion,
including a field-dependent wave function renormalization $Z(\phi)$.
In this paper, this level of approximation has
been analyzed only for the two-dimensional nonlinear sigma model, as in this case it is the first nontrivial order of the derivative expansion.
It is also known  that in the case of quantum critical points
the convergence of this expansion
requires an increasingly accurate tuning of $a$.
For the three-dimensional Wilson-Fisher
	fixed point,  
	this tuning process is expected to converge to optimal
	values within the range $0.5<a<1$~\cite{Balog:2019rrg}.
	Hence, it appears very unlikely that at the special
	point $a\to0$ the derivative expansion might be convergent.

We should mention however, that the amplitude  $a$
is only one of an infinite series of free parameters
within the regulator $R_k$. 
In this work we have not allowed for such residual freedom,
having fixed the regulator to a piecewise linear form.
This choice has been justified as follows.
In some circumstances, 
depending on the theory (or approximation) under study,
as well as on the number of Euclidean dimensions $d$,
the argument of the momentum integral might be
non-integrable in the $a\to0$ limit.
Nonetheless the integral might allow for a finite
$a\to0$ limit, i.e.~the limit and the integral
cannot be exchanged.
Whenever this happens, one must first clearly define the
momentum integrals by choosing a specific shape
function and when applicable a unique analytic continuation,
and then investigate the possible behavior of these integrals
in the parametric $a\to0$ limit.
In all other cases, namely when the $a\to0$ limit
can be brought inside the momentum integrals,
one can easily generalize the discussion to 
arbitrary shape functions $r_1$, as done
in Sec.~\ref{sec:vanishingVScs}. 
Still, optimization criteria over the remaining parameters might be 
essential to obtain accurate results in the
vanishing-regulator limit.
It might also be possible to take advantage of these
additional parameters, with their associated
free limiting behavior, to construct alternative
flow equations resulting from the vanishing-regulator
limit. For instance, in the so-called
LPA${}'$ truncation, 
this kind of additional
freedom allowed
to construct a one-parameter family
of \MSbar -like schemes within the 
FRG~\cite{BPZ}.

 Indeed, as we explained in Sec.~\ref{sec:intro} 
	the limit of vanishing regulator
shares several features with the more specific
case of the \MSbar -like pseudo-regulators 
discussed in Ref.~\cite{BPZ}.
In that reference, and in particular in Sec.~VI, we observed that 
the best way of capturing the effect of quantum/statistical
fluctuations beyond one loop is not adopting
the derivative expansion, but rather accounting for the 
momentum dependence of vertices as in a vertex expansion.
Because of their similarities, it is reasonable to expect that 
this behavior of \MSbar -like pseudo-regulators 
against the choice of truncation scheme
might be shared by the larger class of vanishing regulators.

In spite of the poor results of the $a\to0$ limit of the LPA for the
benchmark case of the Wilson-Fisher fixed point, we think that
this limit may be useful in simple
approximations, in problems where a symmetry is broken by the regulator.
As a first example we have discussed the $O(N+1)$-nonlinear sigma model,
in a formulation where the regulator breaks the global symmetry
to $O(N)$.
In this case we have shown that in the limit of vanishing regulator
the beta functions converge to those of the $O(N+1)$-symmetric theory.

We have then considered the shift symmetry arising in the
background field treatment of a scalar theory.
When this symmetry is broken by the regulator,
this can either generate unphysical fixed points or, 
what is worse,
destroy a physical fixed point.
We have verified that the Ward identities of the shift symmetry
are restored in the limit of vanishing regulator,
and that all the unphysical features of the flow disappear
when $a$ becomes sufficiently small.

It is important to stress the difference between this
logic and the following one that is sometimes found in the FRG literature:
the RG flow equations are solved first (and independently
of the Ward identities) for a parametric family
of regulators; then the latter parameters are tuned such that 
the violation of some finite-dimensional subset of the Ward identities is minimized.
This procedure, 
when applied to the parameter $a$
of \Eqref{eq:generaldefofa}, 
typically results in some nonvanishing value which
is close to the value maximizing the rate of convergence
of the chosen truncation scheme ($a\sim 1$).  This approach 
has been studied for instance in the case of
conformal symmetry~\cite{Balog:2020fyt}.
 In this reference the Ward identities
	for special conformal transformation,
	either in their quantum 
	or classical form
	(i.e.~regulator dependent or independent respectively),
	are not solved as functional 
	constraints.~\footnote{
	The truncated modified Ward identity is cast in the form
    $f(\tilde{\phi})=0$, for a certain function $f$.
    This equation is not fulfilled, for 
    any value of $a$. However it is possible to tune $a$ such that
the function $f$ is minimized in an almost $\tilde\phi$-independent sense.}

    By contrast, in the studies we presented 
        in Secs.~\ref{sec:ON+1} and~\ref{sec:background},
    the ans\"atze for the EAA 
    included exact solutions of the classical
    Ward identites for $O(N+1)$ and shift symmetry
    respectively,
    which  are
    easy to solve independently from the RG equations.
    It is thus not surprising that the symmetry breaking 
    induced by the RG flow is minimized for $a\to0$.
    In fact, one might
    expect that 
    the quantum Ward identities  
    reduce to their classical counterparts when $a\to0$.
  	Thus, because of the different strategy 
  	followed in the choice of the initial 
  	ansatz for the EAA,
  	the authors of 
  	Ref.~\cite{Balog:2020fyt} could only minimize 
  	the unavoidable symmetry breaking,
  	whereas in this work we could tune it 
  	to zero by taking the limit of 
  	vanishing regulators.

It is interesting that a study similar to the one of
Ref.~\cite{Balog:2020fyt} was performed in 
Ref.~\cite{Sonoda:2011qd}, where the symmetry
expected to emerge at the RG fixed point
is supersymmetry rather than
conformal symmetry. In this latter work
the ansatz for the EAA does indeed include
an exact solution of the classical supersymmetric Ward identity.
The minimization of the breaking of supersymmetry at the fixed point
by means of the optimization of the regulator was also studied,
but unfortunately the limit of vanishing regulator was not
within the parametric space considered in this reference.
In fact, we expect  the application
of the vanishing-regulator limit to supersymmetric models
to be interesting and useful.

The main motivation of this work was the hope that
vanishing regulators, or perhaps just 
``sufficiently small regulators'',
may be useful 
also in the application of the FRG to gauge theories
and gravity,
where the background field method is almost always 
 adopted.
Our results suggest that this may be possible,
but that the usefulness of this idea may be restricted
to the simplest truncations.

\section*{Acknowledgments}

This project has received funding from the European Union’s Horizon 2020 research and innovation programme under the Marie Skłodowska-Curie grant agreement No 754496.



\begin{appendix}

\section{Flow equations at the order \texorpdfstring{$\boldsymbol{O(\partial^2)}$}{2} of
the	derivative expansion}
\label{app:floweqDE}

We introduce the following notations
\begin{subequations}
\begin{align}
G_0 &= \left(Z_k(\rho) q^2 + R_k(q^2) + U_k'(\rho) \right)^{-1} ,
\\
G_1 &= \left(\tilde Z_k(\rho) q^2 + R_k(q^2) + U_k'(\rho)+2\rho U_k''(\rho)\right)^{-1} ,
\end{align}
for the Goldstone-bosons and radial-mode propagators.
\end{subequations}
The flow equations for $U_k$ and $\tilde Z_k$,
which is defined in \Eqref{eq:tildeZ}, are
\begin{widetext}
	\begin{align}
		\partial_t U_k &=
		\frac{\left( Q_{\frac{d}{2}}\left[ G_1 \partial_t  R_k \right]
			+ (N-1)Q_{\frac{d}{2}}\left[ G_0 \partial_t  R_k \right] \right)}{2(4\pi)^{d/2}} ,
		\label{eq:DEbetaUQ}
		\\
		\partial_t  \tilde Z_k  &=
		- \frac{\left(\tilde Z_k'+2\rho \tilde Z_k''\right)}{2(4\pi)^{d/2}} Q_{\frac{d}{2}}\left[ G_1^2 \partial_t  R_k \right]
		-(N-1) \frac{\left(  Z_k'+\rho Y_k'\right)}{2(4\pi)^{d/2}} Q_{\frac{d}{2}}\left[ G_0^2 \partial_t  R_k \right]
		\nonumber\\
		&
		+\frac{2 \rho \left( \tilde Z_k'\right)^2}{(4\pi)^{d/2}} \left[
		\frac{2d+1}{2}  Q_{\frac{d}{2}+1}\left[ G_1^3 \partial_t  R_k \right]
		+ \frac{(d+2)(d+4)}{4}\left( Q_{\frac{d}{2}+2}\left[ G_1^2 G_1' \partial_t R_k\right]  +
		Q_{\frac{d}{2}+3}\left[ G_1^2 G_1'' \partial_t  R_k\right]
		\right)\right]
		\nonumber\\
		& 
		+ \frac{2 \rho\left( 3 U_k''+2 \rho U_k''' \right)^2}{(4\pi)^{d/2}} \left( 
		Q_{\frac{d}{2}}\left[ G_1^2 G_1'\partial_t  R_k \right]
		+Q_{\frac{d}{2}+1}\left[ G_1^2 G_1'' \partial_t R_k \right]
		\right)
		\nonumber\\
		&
		+\frac{2 \rho \tilde  Z_k' \left( 3 U_k''+2 \rho U_k'''\right)}{(4\pi)^{d/2}} \left[
		\left( d+2 \right) \left( Q_{\frac{d}{2}+1}\left[ G_1^2 G_1' \partial_t  R_k\right] +
		Q_{\frac{d}{2}+2}\left[ G_1^2 G_1'' \partial_t  R_k\right] \right)
		+2 Q_{\frac{d}{2}}\left[ G_1^3 \partial_t  R_k\right]\right]
		\nonumber\\
		& 
		+ (N-1)\frac{ \rho Y_k }{(4\pi)^{d/2}} \left( 
		2U''_k Q_{\frac{d}{2}}\left[ G_0^3\partial_t  R_k \right]
		+d Z'_k Q_{\frac{d}{2}+1}\left[ G_0^3 \partial_t R_k \right]
		\right)\nonumber\\
		&
		+(N-1)\frac{2 \rho \left(  Z_k'\right)^2}{(4\pi)^{d/2}} \left[
		\frac{(d+2)(d+4)}{4} \left( Q_{\frac{d}{2}+2}\left[ G_0^2 G_0' \partial_t R_k\right] \!+\!
		Q_{\frac{d}{2}+3}\left[ G_0^2 G_0'' \partial_t  R_k\right]
		\right) 
		+\frac{1}{2} Q_{\frac{d}{2}+1}\left[ G_0^3 \partial_t  R_k \right]\right]
		\nonumber\\
		&
		+ (N-1)\frac{2 \rho\left( U_k'' \right)^2}{(4\pi)^{d/2}} \left( 
		Q_{\frac{d}{2}}\left[ G_0^2 G_0'\partial_t  R_k \right]
		+Q_{\frac{d}{2}+1}\left[ G_0^2 G_0'' \partial_t R_k \right]
		\right)
		\nonumber\\
		&
		+(N-1)\frac{2 \rho  Z_k' U_k''}{(4\pi)^{d/2}}
		\left( d+2 \right) \left( Q_{\frac{d}{2}+1}\left[ G_0^2 G_0' \partial_t  R_k\right] +
		Q_{\frac{d}{2}+2}\left[ G_0^2 G_0'' \partial_t  R_k\right] \right) \ .
		\label{eq:DEbetaZQ}
	\end{align}
Using the regulator $R_k= a Z_k (k^2-z)\theta(k^2-z)$~\footnote{We defined $Z_k = Z_k(\rho=0)$ .} we have
\begin{align}
Q_{n}\left[ G^\ell \partial_t  R_k \right] =& \
\frac{1}{\Gamma(n)} \left( (2-\eta_k) k^2 \, q_{n,\ell}\left(a, \omega , \zeta \right) +\eta_k\, q_{n+1,\ell}\left(a, \omega , \zeta \right) \right) \ ,
\\
Q_{n}\left[ G^\ell G' \partial_t  R_k \right] =& \
-\frac{\left( \zeta-a Z_k \right)}{\Gamma(n)} \left( (2-\eta_k) k^2\, q_{n,\ell+2}\left(a, \omega , \zeta \right) +\eta_k\, q_{n+1,\ell+2}\left(a, \omega , \zeta \right) \right) \ ,
\\
Q_{n}\left[ G^\ell G'' \partial_t  R_k \right] =& \
\frac{2\left( \zeta-a Z_k \right)^2}{\Gamma(n)} \left( (2-\eta_k) k^2\, q_{n,\ell+3}\left(a, \omega , \zeta \right) +\eta_k\, q_{n+1,\ell+3}\left(a, \omega , \zeta \right) \right)\nonumber\\
&\ -\frac{a^2\,Z_k^2 \, k^{2(n-\ell-2)}}{\Gamma(n)\left( \zeta + \omega/k^2 \right)^{\ell+2}} \ ,
\end{align}
\end{widetext}
where
\begin{align}
G &= \left( \zeta z + R_k + \omega \right)^{-1}\ ,
\nn
q_{n,\ell}\left(a, \omega , \zeta \right) &= \frac{1}{n}
\frac{a\, Z_k\, k^{2(n-\ell)}}{\left( a Z_k + \omega /k^2 \right)^\ell} 
\\
&\  \times\,\, _2F_1\!\left(\ell,n,n+1;
\frac{a Z_k-\zeta}{a Z_k + \omega /k^2 }\right) ,
\end{align}
and $G$ can be $G_0$ or $G_1$ depending on the choice of $\zeta$ and $\omega$, in particular
\begin{align}
& G=G_0 \hspace{.5cm}\mbox{if} \hspace{.5cm}
\begin{cases} \omega = U_k'(\rho) \ , \\
\zeta = Z_k(\rho) \ , \end{cases}
\\
& G=G_1 \hspace{.5cm}\mbox{if} \hspace{.5cm} 
\begin{cases} \omega = U_k'(\rho)+2\rho U_k''(\rho) \ ,\\
\zeta = \tilde Z_k(\rho) \ . \end{cases}
\end{align}
For the constant regulator \eqref{eq:CSreg} one finds
\begin{align}
&Q_{n}\left[ G^\ell \partial_t  R_k \right] =
k^{2(n-\ell+1)}Z_k(2-\eta_k) 
\nonumber\\
&\ \times
\frac{\Gamma(\ell-n)}{\Gamma(n)\Gamma(\ell)}
\zeta^{-n} 
\left(Z_k+\omega/k^2\right)^{n-\ell} \ ,
\\
&Q_{n}\left[ G^\ell G' \partial_t  R_k \right] =
-\zeta \,Q_{n}\left[ G^{\ell+2} \partial_t  R_k \right]  ,
\\
&Q_{n}\left[ G^\ell G'' \partial_t  R_k \right]  =
2 \zeta^2 \, Q_{n}\left[ G^{\ell+3} \partial_t  R_k \right]  .
\end{align}

\section{Master equation for the nonlinear 
	\texorpdfstring{$\boldsymbol{O( N\! +\! 1)}$}{O(N+1)} model}
\label{app:masterON+1}

In Sec.~\ref{sec:ON+1} the use of a vanishing regulator for the two dimensional nonlinear
$O(N+1)$ model has been discussed.
We have observed that a nonvanishing potential term of the form 
(\ref{eq:Uofh}) is not preserved by the flow equation in the $a\to0$ limit.
In this section we provide more details about the contraints on a general local potential
$U_k(\rho,H)$. Here we show how the non-compatibility of the ansatz
(\ref{eq:Uofh}) with the flow equation is encoded in the modified master equation
for the $O(N+1)/O(N)$ symmetry.

Our starting point is indeed the following
modified master equation 
\begin{equation}
	\frac{\delta \Gamma_k}{\delta \phi^a}\frac{\delta \Gamma_k}{\delta H} + H \, \phi_a =
	\mathrm{Tr} \left\{\! R_k\! \left( \Gamma_k^{(2)} + R_k \right)^{-1}_{ab} \frac{\delta^2 \Gamma_k}{\delta H \delta \phi^b} \right\}.
	\label{eq:exactmastereq}
\end{equation}
This identity, which differs from the 
standard master equation for a nonvanishing rhs,
can be derived for instance from a functional integral representation of $\Gamma_k$,
in presence of a linear source term of the form
(\ref{eq:Uofh}) in the bare action,
by performing a change of integration variable 
corresponding to a $O(N+1)/O(N)$ infinitesimal 
transformation.
It is straightforward to prove that this 
functional identity is compatible with the exact RG
flow equation \cite{Vacca}, meaning that it defines
an RG-invariant hypersurface in theory space.
However, truncations of the theory space often
spoil this property, such that the truncated
master equation 
becomes an additional requirement on the RG flow,
to be enforced at every $k$.

Whenever the regularization preserves the (unmodified)  nonlinear
$O(N+1)/O(N)$ symmetry, the  one-loop regulator-dependent term on the rhs~of \Eqref{eq:exactmastereq} vanishes identically.
The modified master equation then reduces to the standard master equation, which is a tree-level
identity. In this case the equation
is straightforward to solve, independently from
and prior to the analysis of 
the RG flow equation. For an introduction to
the role
played by this identity in the construction of a
renormalized perturbation theory in two dimensions
see for instance Ref.~\cite{ZinnJustin:2007zz}.
Before analysing in details the shape that
this constraint takes for vanishing $a$,
one can already apply its form of \Eqref{eq:exactmastereq}
to the truncation we assumed in Sec.~\ref{sec:ON+1}.
There we took $U=0$ and $H=0$.
This combination trivially fulfills the modified master equation.
It should however be noted that \Eqref{eq:exactmastereq}
represents the constraint of nonlinear $O(N+1)/O(N)$ symmetry
only in the theory space of generic functionals of $\phi^a$ and $H$.
If a nonvanishing $H$ is never introduced in the effective action,
i.e.~on the subspace where $H=0$, there nevertheless is a 
functional constraint encoding the nonlinear $O(N+1)/O(N)$ symmetry,
and it can be obtained from \Eqref{eq:exactmastereq} by replacing derivatives involving $H$
with the expectation values of the corresponding composite operators.
The analysis of this kind of modified master equation is therefore highly nontrivial and
will not be addressed in this work.

We then address the constraints that \Eqref{eq:exactmastereq}
imposes on a truncation similar to the
one in \Eqref{eq:GammaNLSM},
but with an arbitrary nonvanishing
$U_k(\rho,H)$.~\footnote{This general ansatz can be
made compatible with the assumed linear $H$ dependence of the bare action, by requiring the linearity of the 
potential at the UV cutoff scale $k=\Lambda$.}
For this truncation 
\Eqref{eq:exactmastereq}
becomes
\begin{align}
&\partial_\rho U_k \, \partial_H U_k + H  =
\partial_\rho \partial_H U_k
\nonumber\\
&\times
\frac{1}{4\pi}
\int_0^\infty \! dz\,
\frac{  R_k(z) } { 
\tilde{Z}_k \, z + R_k(z) + \partial_\rho U_k + 2\rho \partial_\rho^2 U_k }\ .
\end{align}
For the Litim regulator the loop integral is 
readily evaluated leading to
\begin{align}
&4\pi\frac{\partial_\rho U_k \, \partial_H U_k + H}{\partial_\rho \partial_H U_k}  =
- \frac{a k^2}{a-\frac{g_k^2}{Z_k}
\tilde{Z}_k}
\nonumber\\
&- \frac{a}{\left(a-\frac{g_k^2}{Z_k}
\tilde{Z}_k\right)^2}\frac{g_k^2}{Z_k}\left(
\tilde{Z}_k k^2+\partial_\rho U_k + 2\rho \partial_\rho^2 U_k
\right)
\nonumber\\
&\times
\log\! \left(\frac{\tilde{Z}_k k^2+
\partial_\rho U_k + 2\rho \partial_\rho^2 U_k}{a\frac{Z_k}{g_k^2}k^2+\partial_\rho U_k + 2\rho \partial_\rho^2 U_k}\right).
\label{eq:masterLPA}
\end{align}
The loop contribution to the modified master equation is a 
nonlinear function of derivatives of $U_k$ up to second 
order. Therefore solving this equation 
for $U_k$ is a difficult task. Even more so, as 
this solution must be required to also obey the 
RG flow equation.
As the LPA projection breaks compatibility of the 
modified master equation with the RG flow equation, 
the latter is an independent nonlinear
second order partial differential equation for $U_k$.
This illustrates the difficulty of dealing with
modified Ward identities in the FRG framework.
For a discussion of these issues in the context of gauge theories, see for instance~\cite{Ellwanger:1995qf,Fischer:2004uk}

Can the limit $a\to 0$  be of any help in solving
this complex problem?
In addressing this question we need
to specify the behavior of the functions $U_k$
and $\tilde{Z}_k$ for $a\to0$. 
For definiteness, we assume the scaling 
\begin{equation}
\partial_\rho U\sim a\ ,\quad\quad
\tilde{Z}_k \sim a^0\ ,\quad\quad
\phi\sim a^0\ .\\
\end{equation}
Considering then \Eqref{eq:masterLPA},
it is natural to assume
\begin{equation}
 H\sim a\ ,
\end{equation}
which allows the linear source term to be interpreted as
being part of the potential.
However, inspection of the rhs of~\Eqref{eq:masterLPA}
reveals that the leading behavior of the one-loop contribution
is in fact $a\log a$.
As a consequence we provide an ansatz encoded in the following 
definitions
\begin{align}
H=&\ a\hat{H}\ ,\\
 U_k(\rho,H)=&\ a\, \hat{U}_k(\rho,\hat{H})-a \log a\, F_0(\rho)
 \nonumber\\
 & -a\log(-\log a)\,  F_1(\rho)\ .
 \label{eq:sepUhat}
\end{align}
Notice that we choose an ansatz with $F_0$ and $F_1$ independent of $H$.
This might lead us to a particular solution of the modified master equation.
The modified master equation then can be projected on three distinct equations,
each showing a different small $a$ asymptotic behavior.
The $O(a\log(a))$, $O(a\log(-\log a))$ and $O(a)$ terms in this equation respectively lead to
\begin{align}
& F_0'(\rho)\hat{U}^{(0,1)}(\rho,H)=
 -\frac{Z_k k^2}{4\pi g^2_k}\frac{\hat{U}^{(1,1)}(\rho,H)}{\tilde{Z}_k(\rho)}\ ,
 \label{eq:masterF0}\\
& F_1'(\rho)\hat{U}^{(0,1)}(\rho,H)=
 -\frac{Z_k k^2}{4\pi g^2_k}\frac{\hat{U}^{(1,1)}(\rho,H)}{\tilde{Z}_k(\rho)}\ ,
 \label{eq:masterF1}\\
& H+\hat{U}^{(0,1)}(\rho,H)\hat{U}^{(1,0)}(\rho,H)=
 -\frac{Z_k k^2}{4\pi g^2_k}
 \nonumber\\
&\times \frac{\hat{U}^{(1,1)}(\rho,H)}{\tilde{Z}_k(\rho)}
 \left(\log\bigg(\frac{\tilde{Z}_k k^2}{F_0'(\rho)+2\rho F_0''(\rho)}\bigg)-1\right) .
 \label{eq:masterUhat}
\end{align}
It is evident how the $a\to0$ limit does not relief the nonlinearity of the modified master equation.
While the first two equations can be straightforwardly solved for $F_0$ and $F_1$, once 
$U(\rho,H)$ is known, the third equation is highly nontrivial.
In fact, Eq.~(\ref{eq:masterF0}) can be replaced inside Eq.~(\ref{eq:masterUhat}) to
obtain a second order nonlinear partial differential equation for $\hat{U}$.
While the construction of the most general solution is a very complex task,
which we expect in general to be possible only by numerical methods,
a particular solution can be found by assuming the ansatz
\begin{equation}
 \hat{U}(\rho,H)=\pm\hat{H}\sqrt{\frac{1}{Z_k}-2\rho}+F_2(\rho)\ .
\end{equation}
This leads to a first order ordinary differential equation for $F_2$ which can be
easily solved. The determination of the corresponding $F_0$ and $F_1$ 
results in the following particular solution 
\begin{align}
 F_{0,1}(\rho)=&\ c_{0,1}+\frac{Z_k k^2\rho}{4\pi}\ ,
 \label{eq:solF01}\\
 F_2(\rho)=&\ c_2-\frac{Z_k k^2\rho}{4\pi}\log\!\left(\frac{g^2}{4\pi}\right)
 \nonumber\\
 & +\frac{k^2}{8\pi}(1-2Z_k\rho)\log(1-2Z_k\rho)\ ,
 \label{eq:solF2}
\end{align}
where $c_{0,1,2}$ are integration constants which can depend on $k$.
Having an analytic formula for a particular solution of the master equation, 
is of course a nice result, which is possible only thanks to the simplifications
brought by the $a\to0$ limit. However, in itself this result is of limited use,
for two main reasons.
First, in general there is no reason to expect that this anstaz be closed under the
RG flow. Given the compatibility of the flow equation with the master equation,
any particular solution is free to flow into the most general solution
during an infinitesimal RG step. Second, in the LPA case even this 
compatibility is lost. As a consequance, the solution of Eqs.~(\ref{eq:sepUhat}), (\ref{eq:solF01}), and (\ref{eq:solF2})
will flow into a potential which does not fulfill the modified master equation.
Therefore, this solution would be  useful only if accompanied by a prescription
for projecting the latter potential back onto
a functional of the same form as the particular solution itself.
We do not explore possible prescriptions for this projection in this work.

\end{appendix}

\newpage


\end{document}